\documentclass[aps,preprint,showpacs ,amssymb, amsmath,pre]{revtex4}
\usepackage{bm}
\usepackage{graphicx}

\begin{document}

\title{Thomson scattering in high intensity regime}

\author{Madalina Boca}
\email{madalina.boca@g.unibuc.ro}
\affiliation{Department of Physics,  University of Bucharest, MG-11, Bucharest-M\u agurele, 077125  Romania}

\author{Andreea Oprea}
\affiliation{Department of Physics,  University of Bucharest, MG-11, Bucharest-M\u agurele, 077125  Romania}

\begin{abstract}
 Within the framework of the classical electrodynamics, we investigate  the scattering of a very  intense laser pulse   on ultrarelativistic electrons. The laser pulse is modeled by a plane wave with finite length. For  a circularly polarized laser pulse, we focus on the angular distribution of the emitted radiation 
 in its dependence on the electron energy for the cases of  head-on and  90 degrees collisions. We investigate the relation between $dW/d\Omega$  and the trajectory followed by the  velocity of the electron during the laser pulse and, for the case of a short laser pulse, we discuss the carrier-envelope phase  effects. We also present an analysis of the polarization of the emitted radiation. We find two scaling laws  allowing  to predict the behaviour of the angular distributions for a broader range of parameters.
\end{abstract}
\maketitle

\section{Introduction}\label{s-intro}
The description of the scattering of  very intense electromagnetic radiation by electrons has become an interesting subject for the theorists after the invention of the laser. When the description within the classical electrodynamics (CED) is valid, the process is usually named non-linear Thomson scattering; the term non-linear Compton scattering is used when the quantum description is required. 

 The first step in the quantitative description of  non-linear Thomson effect is the solving of relativistic equations of motions for the electron in the laser field, then the solution is plugged into the well known expression of the Lienard-Wiechert potentials \cite{Jackson}.  Many of the existing calculations \cite{SS,esa, esa2,Faisal1, Faisal2,yu} refer to the case of a monochromatic laser field, where   the double differential spectrum $d^2W/d\Omega d\omega$ of the emitted radiation consists in an infinite series of equidistant lines whose positions depend on the laser intensity, electron momentum and observation direction.  Calculations for the case of a finite plane-wave laser pulse were also  performed  \cite{as,lee2}; Krafft {\it et al.} \cite{krafft1,krafft2} gave  an analysis of polarization of the emitted radiation for a particular geometry. Results of numerical calculations for more realistic models of a focused laser pulse \cite{gao,brown,lan-focus,lee,heinzl2} or based on the solutions of the Dirac-Lorentz equation \cite{dirac} were also published. 

 For the quantum description of intense   radiation scattering on free electrons  and published results based on it, we mention two recent review papers \cite{sal,ehl}.
Of particular interest, especially for determining the validity limit of the CED formalism, is the relation between the classical and quantum results.  For a  monochromatic field, Goreslavskii {\it et al.} \cite{GRSL} proved that  the differential cross section of the Thomson scattering can be obtained as the classical limit $(\hbar\rightarrow0)$ of the corresponding quantum expression and  they also discussed   validity conditions  for the classical approach. More recently,  Heinzl {\it et al.} \cite{heinzl2} analyzed  the relation between the Thomson and Compton scattering for the terms corresponding to the  absorption  of a fixed number a photons in the cross sections, an approach possible in the monochromatic case only.  In the case of a plane wave pulse,   Boca and Florescu \cite{BF-u}   derived a new analytic expression for  the  classical double differential spectrum $d^2W/d\Omega d\omega$ of the emitted radiation  by taking the classical limit of the quantum formula \cite{BF} and confirmed by numerical calculations the existent estimations of the  limit of validity  of the classical approach.

The interest in the study of the radiation scattering resides in the potential applications of the process for the domain in the continuous development of laser technology. The possible application of the Thomson/Compton process as a  source of very short pulses, from picosecond to attosecond and even zeptosecond domain,  was considered in several papers \cite{esa,lan-focus,as,zhang,chung,zepto}. Another application is related to the determination of the relative phase between the carrier and the envelope (CEP) for the case of a few cycle laser pulses using Thomson/Compton scattering measurements \cite{gao, lan2, keitel}. 

Experimental work devoted to non-linear effects in radiation scattering from free electrons  includes the observation of  non-linear Thomson effect in the case of non-relativistic electrons  \cite{enri,kumita,babzien} and   the detection of non-linear  Compton effect, reported  \cite{SLAC} in an experiment of collision between GeV electrons and a laser pulse of intensity of the order of atomic unit $I_0\sim10^{16}\;{\mathrm{W/cm^2}}$. A recent proposal for an experiment \cite{ELI}, made in connection to the envisaged construction of the ELI-NP facility, suggests the experimental investigation of scattering of an ultraintense laser pulse $(I_{\mathrm L}\ge 10^{22}\;{\mathrm{W/cm^2}})$ on MeV electrons. In this regime the process it still within  or at  the boundary of the CED formalism  domain of  validity; new features are expected to appear as a consequence of the balance between the effect of the very intense laser field, on the one hand, and the large energy of the electron,  on the other. Theoretical calculations referring to this range of intensities and energies \cite{heinzl2,ELI} are focused mainly on the energetic spectrum of the emitted radiation. The purpose of our  paper is to give a first general overview of the angular distribution $dW/d\Omega$ which could be observed in such an experiment.  

We start  Sect. \ref{s-theory} with the analytical expressions of the trajectory, velocity and acceleration for a charged particle in interaction with an electromagnetic pulse with fixed propagation direction but arbitrary length and shape. Next, we review the main  equations of the classical theory of radiation scattering and discuss a high energy approximation of the exact CED formula; we also give an exact and an approximate expression for the angular distribution corresponding to two particular state of polarization of the emitted radiation. Section \ref{s-numerical} contains numerical examples. For a circularly polarized laser pulse with fixed intensity $I_{\mathrm L}\approx3\times10^{21}\;{\mathrm{W/cm^2}}$, we present graphs of $dW/d\Omega$ for two collision geometries and  different electron energies, and we illustrate the relation between the angular distribution and the trajectory followed by the velocity of the electron during the interaction with the laser pulse; two scaling laws are briefly discussed. For the case of a short pulse, we study the CEP effect on the angular distribution of the emitted radiation. Finally, we present angular distributions for the radiation emitted with given polarization state and we compare them with the analogous results based on the high-energy approximation  we have proposed in Sect. \ref{s-theory}. Section \ref{s-conclusions} contains our conclusions.
\section{Theory}\label{s-theory}
\subsection{The electron trajectory}\label{ss-el_traj}
The motion of a charged particle in an electromagnetic plane-wave is a textbook problem \cite{LL}; we briefly present here the results which will be used in the following.  We consider a plane wave electromagnetic pulse with the propagation direction ${\bf n}_{\mathrm L}$ chosen along the third axis of the reference frame, ${\bf n}_{\mathrm L}\equiv {\bf e}_z$  described by a vector potential ${\bf A}$ orthogonal to the propagation direction, and depending on time and coordinates only through the combination $\chi\equiv ct-{\bf n}_{\mathrm L}\cdot{\bf r}$, where $c$ is the velocity of light. It is convenient to introduce the four-vector $n_{\mathrm L}\equiv (1, {\bf n}_{\mathrm L})$; with this notation the argument $\chi$ of the vector potential can be expressed as the four-product $\chi=x\cdot n_{\mathrm L}$,  with $x$  the coordinate four-vector $x\equiv (ct, {\bf r})$.  Although the trajectory can not be written as an {\it explicit} function of time,  it is possible to write the trajectory, velocity and acceleration as {\it explicit} functions of $\chi$. We shall describe the case of an electron of mass $m$ and electric charge $e<0$ which is placed in the origin of the reference frame at a moment $t_0$ when the laser pulse is far from the electron, and has the initial momentum $p_0\equiv(E_0/c,{\bf p}_0)$; we denote by ${\bm{\beta}}_0$ the initial velocity, measured in units of $c$
\begin{equation}
{\bm{\beta}}_0\equiv\frac{{\bf v}_0}{c}=\frac{{\bf p}_0}{\sqrt{(mc)^2+{\bf p}_0^2}},\label{def-beta_0}
\end{equation}
and introduce the Lorentz factor
\begin{equation}
\gamma=\frac1{\sqrt{1-\beta_0^2}}.\label{def-gamma}
\end{equation}
 For any vector we shall use the index $\perp$ to indicate the components orthogonal to the propagation direction ${\bf n}_{\mathrm L}$, and the component along ${\bf n}_{\mathrm L}$ will carry the index $z$; also, the following notations will be used:
\begin{equation}
e^2{\cal A}^2(\chi)=e^2{\bf A}^2(\chi)-2e{\bf A}(\chi)\cdot{\bf p}_{0\perp},\label{cal_A}
\end{equation}
\begin{equation}
F(\chi)=\frac{e^2{\cal A}^2(\chi)}{2(n_{\mathrm L}\cdot p_0)^2}+\frac{p_{0z}}{n_{\mathrm L}\cdot p_0}.\label{F}
\end{equation}
With the previous notations one obtains, after a change of variable from $t$ to $\chi$ in  the relativistic equations of motion for the electron in the electromagnetic field ${\bf A}(\chi)$,
\begin{equation}
{\bf r}_{\perp}(\chi)=\int\limits_{ct_0}^{\chi}d\rho\frac{{\bf p}_{0\perp}-e{\bf A}(\rho)}{n_{\mathrm L}\cdot p_0},\qquad z(\chi)=\int\limits_{ct_0}^{\chi}d\rho F(\rho);\label{eq_traj}
\end{equation}
\begin{equation}
{\bm{\beta}}_{\perp}(\chi)\equiv\frac{\dot{\bf r}_{\perp}}{c}=\frac{{\bf p}_{0\perp}-e{\bf A}(\chi)}{n_{\mathrm L}\cdot p_0}\frac1{1+F(\chi)},\qquad\beta_z(\chi)\equiv\frac{\dot z}{c}=\frac{F(\chi)}{1+F(\chi)};\label{eq-beta}
\end{equation}
\begin{equation}
\dot{\bm{\beta}}_{\perp}(\chi)\equiv\frac{\ddot{\bf r}_{\perp}}{c}=-\frac1{(1+F(\chi))^2}\frac1{n_{\mathrm L}\cdot p_0}\left[e\frac{d{\bf A}(\chi)}{d\chi}+\frac{dF(\chi)}{d\chi}\frac{{\bf p}_{0\perp}-e{\bf A}(\chi)}{1+F(\chi)}\right],\label{eq-dotbeta-perp}
\end{equation}
\begin{equation}
\dot\beta_z\equiv\frac{\ddot z}{c}=\frac1{(1+F(\chi))^3}\frac{dF(\chi)}{d\chi}.\label{eq-dotbeta-z}
\end{equation}
In the previous equation the symbols dot and doubledot  indicate the  first and, respectively, second order derivatives with respect to the time $t$;  the expression of $\dot\chi$,
\begin{equation}
\dot\chi\equiv\frac{d\chi}{dt}=c(1-\beta_z)=\frac{c}{1+F(\chi)},\label{dotchi}
\end{equation}
 will be also useful. 
\subsection{Angular distribution of scattered radiation}
In classical electrodynamics formalism the angular distribution of the radiation emitted by an accelerated charge $e$ is given by \cite{Jackson}
\begin{equation}
\frac{dW}{d\Omega}=\frac{e_0^2}{4\pi c}\int\limits_{-\infty}^{\infty}dt\frac1{\kappa^5}|{\bf n}\times[({\bf n}-{\bm{\beta}})\times\dot{\bm{\beta}}]|^2,\qquad e_0^2=\frac{e^2}{4\pi\epsilon_0}\label{dW-t}
\end{equation}
where ${\bf n}$ is the unit vector of the observation direction, characterized by the angles $\theta$ and $\phi$, ${\bm{\beta}}$ and $\dot{\bm{\beta}}$ are, respectively the particle velocity and acceleration, and the factor  $\kappa$ in the denominator is
\begin{equation}
\kappa=1-{\bf n}\cdot{\bm{\beta}}.\label{kappa}
\end{equation}
For the case of an electron accelerated in a plane wave electromagnetic field, since the electron velocity and acceleration can be written  as explicit functions of $\chi$, it is convenient to perform a change of variable from $t$ to $\chi=n\cdot x$ in the integral (\ref{dW-t}).  Using the derivative of $\chi$, given by Eq. (\ref{dotchi}), we obtain the alternative expression
\begin{equation}
\frac{dW}{d\Omega}=\frac{e_0^2}{4\pi c^2}\int\limits_{-\infty}^{\infty}d\chi\frac{1+F(\chi)}{(1-{\bf n}\cdot{\bm{\beta}}(\chi))^5}|{\bf n}\times[({\bf n}-{\bm{\beta}}(\chi))\times\dot{\bm{\beta}}(\chi)]|^2;\label{dW-chi}
\end{equation}
for simplicity in the following we shall use the notation
\begin{equation}
{\bf w}(\chi)\equiv {\bf n}\times[({\bf n}-{\bm{\beta}}(\chi))\times\dot{\bm{\beta}}(\chi)].\label{def-w}
\end{equation}
For each direction of observation ${\bf n}$, $dW/d\Omega$ can be decomposed in two components describing the contributions of two polarization states of the emitted radiation. Choosing two unit vectors orthogonal to each other and to the observation direction ${\bf n}$,
\begin{equation}
{\bm{\epsilon}}_1\cdot{\bm{\epsilon}}_2=0,\qquad{\bm{\epsilon}}_i\cdot{\bf n}=0,\quad {\bm{\epsilon}}_i^2=1\,,\qquad i=1,2\,,
\end{equation}
the components of ${\bf w}(\chi)$ along the polarization vectors ${\bm{\epsilon}}_i$ are 
\begin{equation}
w_i(\chi)\equiv{\bf w}(\chi)\cdot{\bm{\epsilon}}_i=-\beta_i(\chi)\dot{\bm{\beta}}(\chi)\cdot{\bf n}-\dot\beta_i(\chi)\kappa\,,\qquad i=1,2\,,
\end{equation}
where $\beta_i$ and $\dot\beta_i$ are the components of the ${\bm{\beta}}$ and, respectively $\dot{\bm{\beta}}$ along ${\bm{\epsilon}}_i$.  We decompose the angular distribution of the emitted radiation  as 
\begin{equation}
\frac{dW}{d\Omega}=\frac{dW_1}{d\Omega}+\frac{dW_2}{d\Omega},\qquad\frac{dW_i}{d\Omega}=\frac{e_0^2}{4\pi c^2}\int\limits_{-\infty}^{\infty}d\chi\frac{1+F(\chi)}{(1-{\bf n}\cdot{\bm{\beta}}(\chi))^5}w_i^2(\chi)\,,\quad i=1,2\,.\label{dW-pol}
\end{equation}
In the numerical examples presented in the next section we will choose the polarization vectors $\bm {\epsilon}_1$   and $\bm {\epsilon}_2$ as
\begin{equation}
{\bm{\epsilon}}_1=\frac{{\bf n}\times{\bf n}_{\mathrm L}}{|{\bf n}\times{\bf n}_{\mathrm L}|},\qquad{\bm{\epsilon}}_2=\frac{{\bf n}\times({\bf n}\times{\bf n}_{\mathrm L})}{|{\bf n}\times {\bf n}_{\mathrm L}|}\label{epsilon}.
\end{equation}
With ${\bf n}_{\mathrm L}$ chosen along $Oz$ and ${\bf n}$ characterized by the polar angles $\theta$ and $\phi$,  we have
\begin{equation}
{\bm{\epsilon}}_1={\bf e}_x\sin\theta\sin\phi-{\bf e}_y\sin\theta\cos\phi,\qquad{\bm{\epsilon}}_2={\bf e}_x\cos\theta\cos\phi+{\bf e}_y\cos\theta\sin\phi-{\bf e}_z\sin\theta\,.\label{def-epsilon}
\end{equation}

Next we present a high energy approximation of the angular distribution formula (\ref{dW-chi}). For the case of an ultrarelativistic particle  ($\beta\lesssim 1$) the emission at any moment is concentrated in a very small cone along the instantaneous  direction of ${\bm{\beta}}$ \cite{Jackson}. For the angular distribution $dW/d\Omega$, it follows that the radiation will be observed practically only in those directions ${\bf n}$ which are in, or close to, the range of directions swept by ${\bm{\beta}}$ during the interaction of the electron with the laser pulse;  this can be quantitatively understood from the presence of the denominator $(1-{\bm{\beta}}\cdot{\bf n})^5$ in the expression (\ref{dW-chi}) of $dW/d\Omega$. Let us assume that, for an ultrarelativistic electron, the observation direction is close to the direction reached by ${\bm{\beta}}$  in (\ref{eq-beta}) for a given value $\chi_0$ of the variable $\chi$
\begin{equation}
{\bf n}\approx\widehat{\bm{\beta}}(\chi_0) \equiv\frac{{\bm{\beta}}(\chi_0)}{\beta(\chi_0)};
\end{equation}
then, for that direction of observation only the values $\chi$ close to $\chi_0$ will contribute to the integral (\ref{dW-chi}).  As a consequence, we can approximate the integrand by replacing in the expression (\ref{def-w}) of  ${\bf w}(\chi)$ the velocity ${\bm{\beta}}(\chi)$ by ${\bm{\beta}}(\chi)\approx\widehat{\bm{\beta}}(\chi_0)\beta(\chi)\approx{\bf n}\beta(\chi)$,
\begin{equation}
{\frac{dW}{d\Omega}\vline}_{\;{\bf n}\approx\widehat{\bm{\beta}}(\chi_0)}\approx\frac{d\widetilde W}{d\Omega}\equiv\frac{e_0^2}{4\pi c^2}\int\limits_{-\infty}^{\infty}d\chi\frac{1+F(\chi)}{(1-{\bf n}\cdot{\bm{\beta}}(\chi))^5}(1-\beta(\chi))^2|{\bf n}\times[{\bf n}\times\dot{\bm{\beta}}(\chi)]|^2.\label{dW-app}
\end{equation}
As for an observation direction far from the range of directions taken by the electron velocity the emission is anyway negligible, we can use in fact the approximate expression $d\widetilde W/d\Omega$ for any ${\bf n}$; in the next section this assumption will be numerically tested. The decomposition in the two polarized components is then simply
\begin{equation}
\frac{d\widetilde W}{d\Omega}=\frac{d\widetilde W_1}{d\Omega}+\frac{d\widetilde W_2}{d\Omega},\qquad\frac{d\widetilde W_i}{d\Omega}=\frac{e_0^2}{4\pi c^2}\int\limits_{-\infty}^{\infty}d\chi\frac{1+F(\chi)}{(1-{\bf n}\cdot{\bm{\beta}}(\chi))^5}(1-\beta(\chi))^2\dot\beta^2_i(\chi)\,,\,i=1,2\,,\label{dW-pol-app}
\end{equation}
 where $\dot\beta_i$ are the two components of the acceleration along the two polarization vectors (\ref{def-epsilon}). For understanding the numerical results presented in Sect. \ref{s-numerical} it is useful to  to have the expression of  $\dot\beta_i$ calculated for ${\bf n}=\hat{\bm{\beta}}$, 
\begin{equation}
{\dot\beta_1\vphantom{\frac12}\vline}_{\,{\bf n}=\hat{\bm{\beta}}}=\frac1{|{\bm{\beta}}\times{\bf{n}}_{\mathrm L}|}(\dot\beta_x\beta_y-\beta_x\dot\beta_y),\quad {\dot\beta_2\vphantom{\frac12}\vline}_{\,{\bf n}=\hat{\bm{\beta}}}=\frac1{|{\bm{\beta}}\times({\bm{\beta}}\times{\bf{n}}_{\mathrm L})|}\left[\frac12\beta_z\frac{d{\bm{\beta}}_{\perp}^2}{dt}-\dot\beta_z{\bm{\beta}}_{\perp}^2\right].\label{dot-beta-app}
\end{equation}

In the end of this section we mention that we have checked that approximating further the expression (\ref{dW-pol-app}) as
\begin{equation}
\frac{d\widetilde W_i}{d\Omega}=\frac{e_0^2}{4\pi c^2}\int\limits_{-\infty}^{\infty}d\chi\frac{1+F(\chi)}{(1-{\bf n}\cdot{\bm{\beta}}(\chi))^3}\dot\beta^2_i(\chi)\,,\qquad\,i=1,2
\end{equation}
leads to  practically identical numerical results;  this second approximation will not be used in the following.

\section{Numerical results}\label{s-numerical}

We  consider  the case of a circularly polarized plane wave laser pulse whose shape is modeled by an envelope with  Gaussian wings and a central region of constant amplitude  and variable length. Such a field, with the propagation direction ${\bf n}_{\mathrm L}$ chosen along the $Oz$ axis of the reference frame, can be described by the vector potential
\begin{equation} 
{\bf A}(\chi)=\frac{A_0}{\sqrt{2}}f(\chi)[{\bf e}_x\sin(k_{\mathrm L}\chi)+{\bf e}_y\cos(k_{\mathrm L}\chi)],\qquad k_{\mathrm L}=\frac{\omega_{\mathrm L}}c\,.\label{def-A}
\end{equation}
We take an envelope $f(\chi)$ given by
\begin{equation}
f(\chi)=\left\{\begin{array}{ll}\exp[-1.386\,{k_{\mathrm L}^2\chi^2}/({4\pi^2\tau^2})],&\chi\le0\\1,&0\le\chi\le N_{\mathrm c}cT_{\mathrm L}\\\exp[-1.386\,{k_{\mathrm L}^2(\chi-N_{\mathrm c}cT)^2}/({4\pi^2\tau^2})],&\chi\ge N_{\mathrm c}cT_{\mathrm L}\end{array}\right.,\qquad T_L=\frac{2\pi}{\omega_{\mathrm L}}\label{def-env}
\end{equation}
With this choice of the envelope the parameter $\tau$ is the full width at half maximum (FWHM) of the Gaussian wings, measured in periods $T_{\mathrm L}$ and $N_{\mathrm c}$ is the length of the flat region, also measured in units of $T_{\mathrm L}$. In all the numerical calculation presented in this section the central frequency is chosen $\omega_{\mathrm L}=0.043\;{\mathrm{au}}$ ($\lambda=1060\;{\mathrm{nm}}$), close to the fundamental frequency of the   Nd:YAG laser.  The maximum intensity of the laser pulse is described by the dimensionless parameter 
\begin{equation}
\eta=\frac{|e|A_0}{mc}\label{def-eta};
\end{equation}
 in the numerical calculations presented here we will choose $\eta=50$ which corresponds to the intensity $I_{\mathrm L}\approx3\times10^{21}\;{\mathrm{W/cm^2}}$; for the electron Lorentz factor will be chosen values $\gamma\in(10,45)$.    At the end of this section we shall present two scaling laws which allows us, using the numerical results presented here, to predict the behaviour of the angular distribution also for different values of the laser field intensity and electron energy.

Before presenting the numerical results, we justify    the validity of the the classical description in the mentioned conditions.

 In the study  of the relation between classical and quantum description  of the radiation scattering  done by Heinzl {\it et al.} \cite{heinzl2} for the case of monochromatic radiation, in which case the spectrum for a fixed observation direction consists in an infinite series of discrete lines,  it was shown that the condition for the $l$-th harmonic to be described correctly in the classical formalism is
\begin{equation}
y_l\equiv 2l\frac{\hbar\omega_{\mathrm L}(n_{\mathrm L}\cdot p_0)}{m^2c^3(1+\eta^2/2)}\ll1.
\end{equation}
In the high intensity limit the number of emitted harmonics is $l_{\mathrm{eff}}\sim\eta^3$ \cite{SS}, \cite{GRSL}, and using the inequality $n_{\mathrm L}\cdot p_0\le2mc\gamma$, in the limit $\eta\gg1$ the above condition  becomes 
\begin{equation}
\widetilde y\equiv \frac{8\eta\gamma\hbar\omega_{\mathrm L}}{mc^2}\ll1.
\end{equation}
as for $\eta=50$ and $\gamma=45$ one obtains $\widetilde y\approx0.04$,   we can conclude that for the cases considered here the classical electrodynamics formalism is valid.

An equivalent criterion for the validity limit of the classical approximation is discussed by Mackenroth {\it et al} \cite{keitel}. For the case of head-on collisions they have considered, they write $\xi\ll1$, with 
\begin{equation}
\xi=\frac{\eta\hbar \omega_{\mathrm L}(E_0+c|{\bf p}_0|)}{m^2c^4}\mathop{\approx}_{\gamma\gg 1}\frac{2\eta\gamma\hbar\omega_{\mathrm L}}{mc^2};
\end{equation}
the authors mention that the physical interpretation of the parameter $\chi$ is the ratio between the amplitude of the electric component of the laser field seen in the rest frame of the the incident electron, ${\cal E}_0=(E_0+c|{\bf p}_0|)\omega_{\mathrm L}\eta/|e|c$, and the quantum electrodynamics critical field ${\cal E}_{\mathrm{cr}}=m^2c^3/|e|\hbar$.

In the following we present the results of our calculations for two scattering geometries: in the first case,  at the initial moment, when the laser pulse is still far from the origin, the electron propagates along the $Oz$ axis, in the opposite sense with respect to the laser pulse (the so-called head-on collision); the second example refers to the 90 degrees geometry, when the initial electron direction is chosen orthogonal to the laser propagation direction, along the $Oy$ axis. In each case we  present the angular distribution of the emitted radiation $dW/d\Omega$ in a color logarithmic scale; the values marked on the colorbar next to each graph represent $\ln(dW/d\Omega)$, with $dW/d\Omega$ expressed in atomic units.

\subsection{The case of head-on geometry}\label{ss-head-on}

\begin{figure}
\includegraphics[scale=0.25]{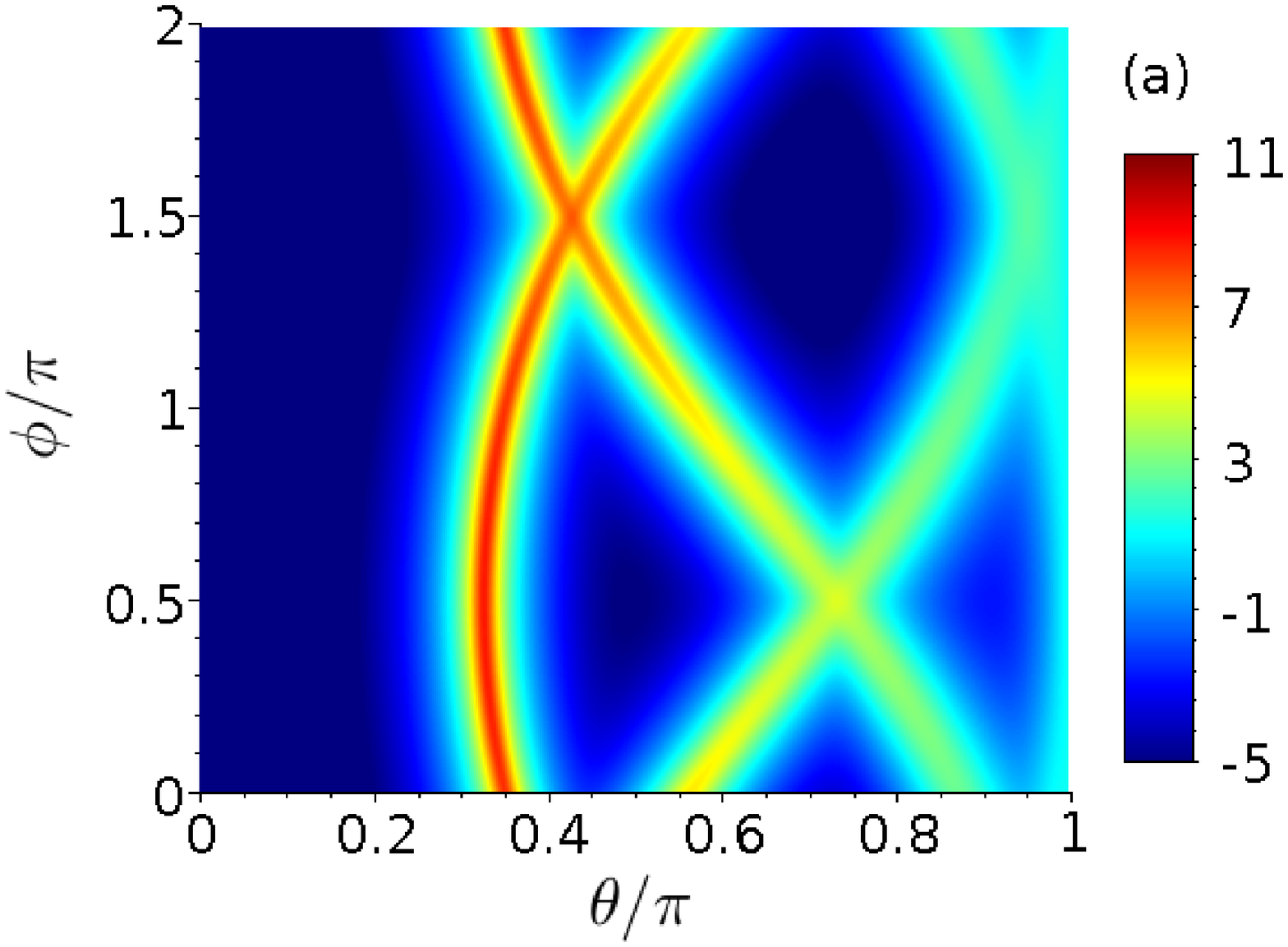} 
\includegraphics[scale=0.25]{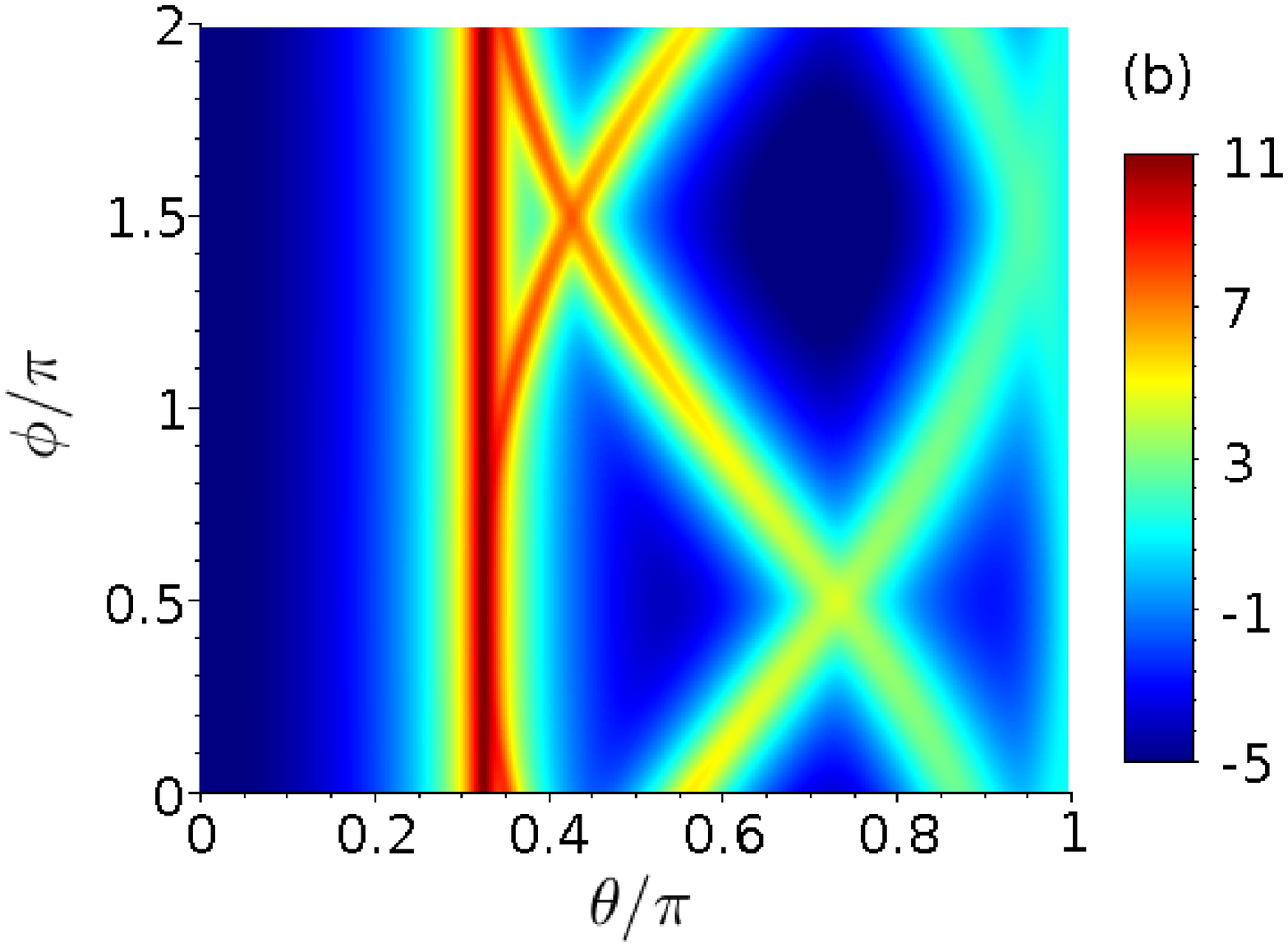}
\includegraphics[scale=0.19]{f1-c.eps}
\caption{(Color online) (a): $\frac{dW}{d\Omega}$ for head on collision, $\tau=1$, $N_{\mathrm c}=0$, $\gamma=10$; (b): the same as (a), except for $N_{\mathrm c}=10$; (c) the trajectory of $\widehat{\bm{\beta}}$ in the plane $(\theta,\phi)$ in the same  conditions as (b)\label{f1}}
\end{figure}

The first case we investigate is when the initial electron propagates in the negative sense of the $Oz$ axis
\begin{equation}
{\bf p}_0=-p_0{\bf n}_{\mathrm L}.\label{head-on}
\end{equation} For the beginning we shall present an analysis of the relation between the angular distribution of the emitted radiation and the direction of electron velocity; as explained already, one can expect to see maxima of emitted radiation in those directions which are reached by ${\bm{\beta}}$  during the interaction with the laser pulse. In Fig. \ref{f1}(a) is represented $dW/d\Omega$ for a laser pulse with $\tau=1$ (FWHM$\approx$3.5 fs) and $N_{\mathrm c}=0$ (i.e. a pulse consisting only in the two wings) and for the electron Lorentz factor $\gamma=10$; Fig. \ref{f1}(b) corresponds to the same conditions, except for the fact that the laser pulse has a flat region of length $N_{\mathrm c}=10$. The two figures are  similar; the contribution of the flat part of the laser pulse in (b) is concentrated in the very intense straight line at $\theta\approx0.32\pi$, the contribution of the wings being identical in the two cases. In Fig. \ref{f1}(c) is represented the trajectory followed by the unit vector of the velocity, $\widehat{\bm{\beta}}$, during the interaction with the pulse.

From the velocity expression (\ref{eq-beta}) evaluated for the vector potential (\ref{def-A}),  (\ref{def-env}),  with the initial condition (\ref{head-on}), it follows that during the constant part of the pulse ${\bm{\beta}}$ moves along a circle of radius
\begin{equation}
R_0=\frac{eA_0}{\sqrt{2}}\frac{n_{\mathrm L}\cdot p_0}{(n_{\mathrm L}p_0)^2+e^2A_0^2/4+p_{0z}(n_{\mathrm L}\cdot p_0)}
\end{equation}
parallel to the plane $Oxy$, at the constant height 
\begin{equation}
Z_0=\frac{e^2A_0^2/4+p_{0z}(n_{\mathrm L}\cdot p_0)}{(n_{\mathrm L}p_0)^2+e^2A_0^2/4+p_{0z}(n_{\mathrm L}\cdot p_0)}.\label{def-Z0}
\end{equation}
The corresponding trajectory of $\widehat{\bm{\beta}}$ in the plane $(\theta,\phi)$ is a straight vertical line located at 
\begin{equation}
\Theta_0=\arccos(Z_0/R_0),
\end{equation}
covered $N_{\mathrm c}$ times during the interaction with the laser. This part of the trajectory was represented by a red line in Fig. \ref{f1}(c) and it leads to the corresponding very bright portion in Fig. \ref{f1}(b). For the following  results we shall not represent the trajectory of $\widehat{\bm{\beta}}$, but we have checked in all cases its  agreement with the shape of the angular distribution.

\begin{figure}
\includegraphics[scale=0.24]{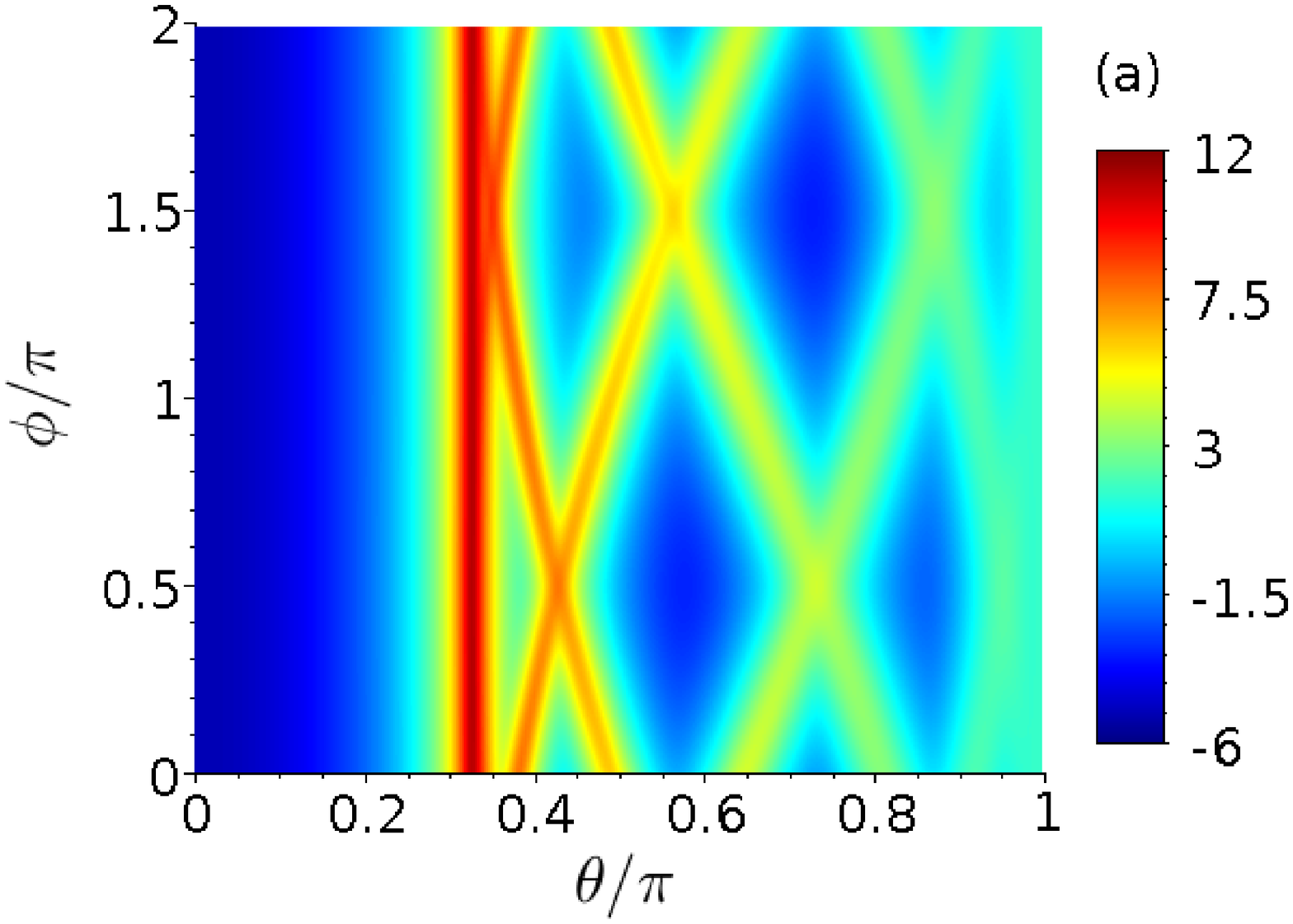}
\includegraphics[scale=0.24]{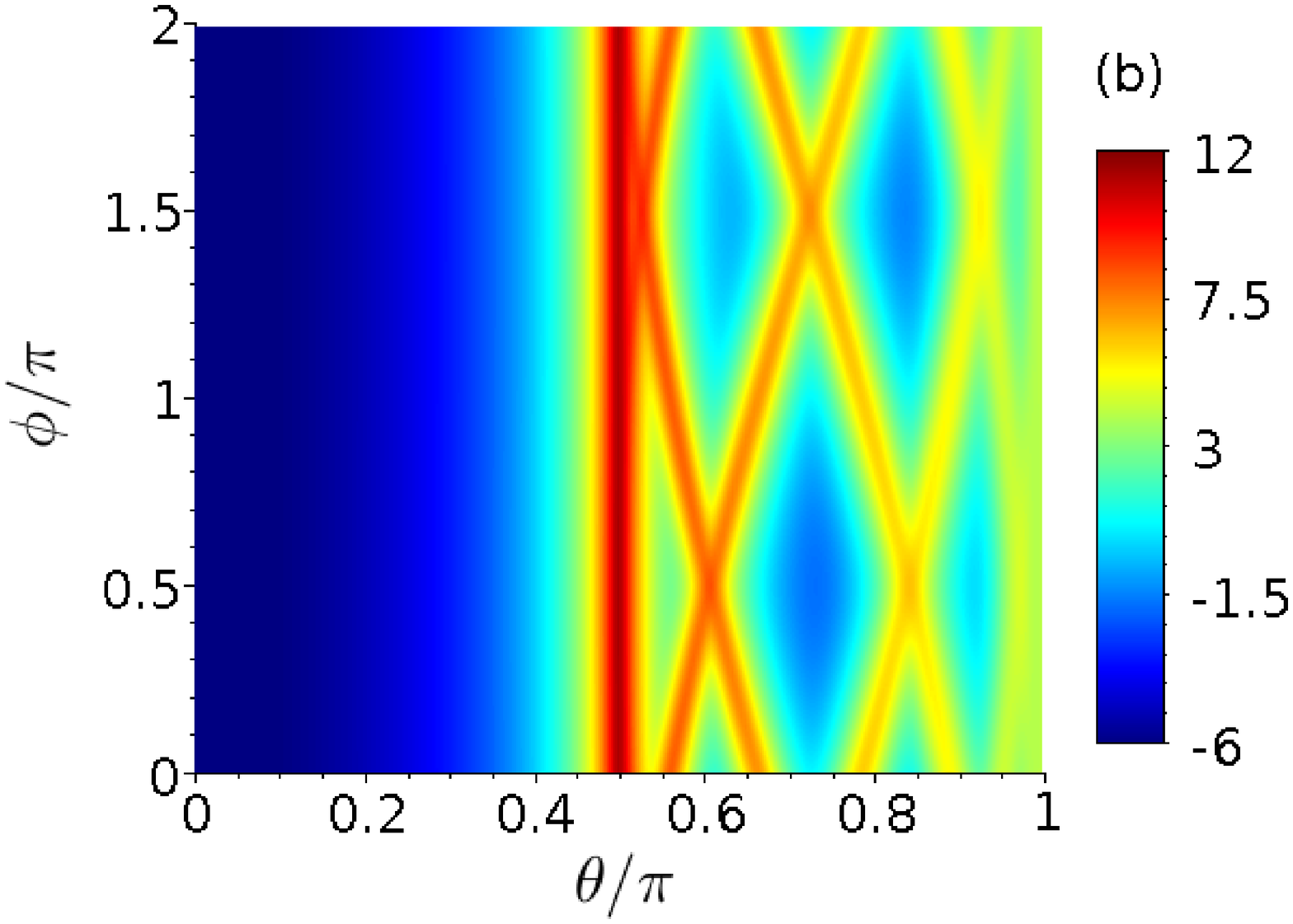}
\includegraphics[scale=0.24]{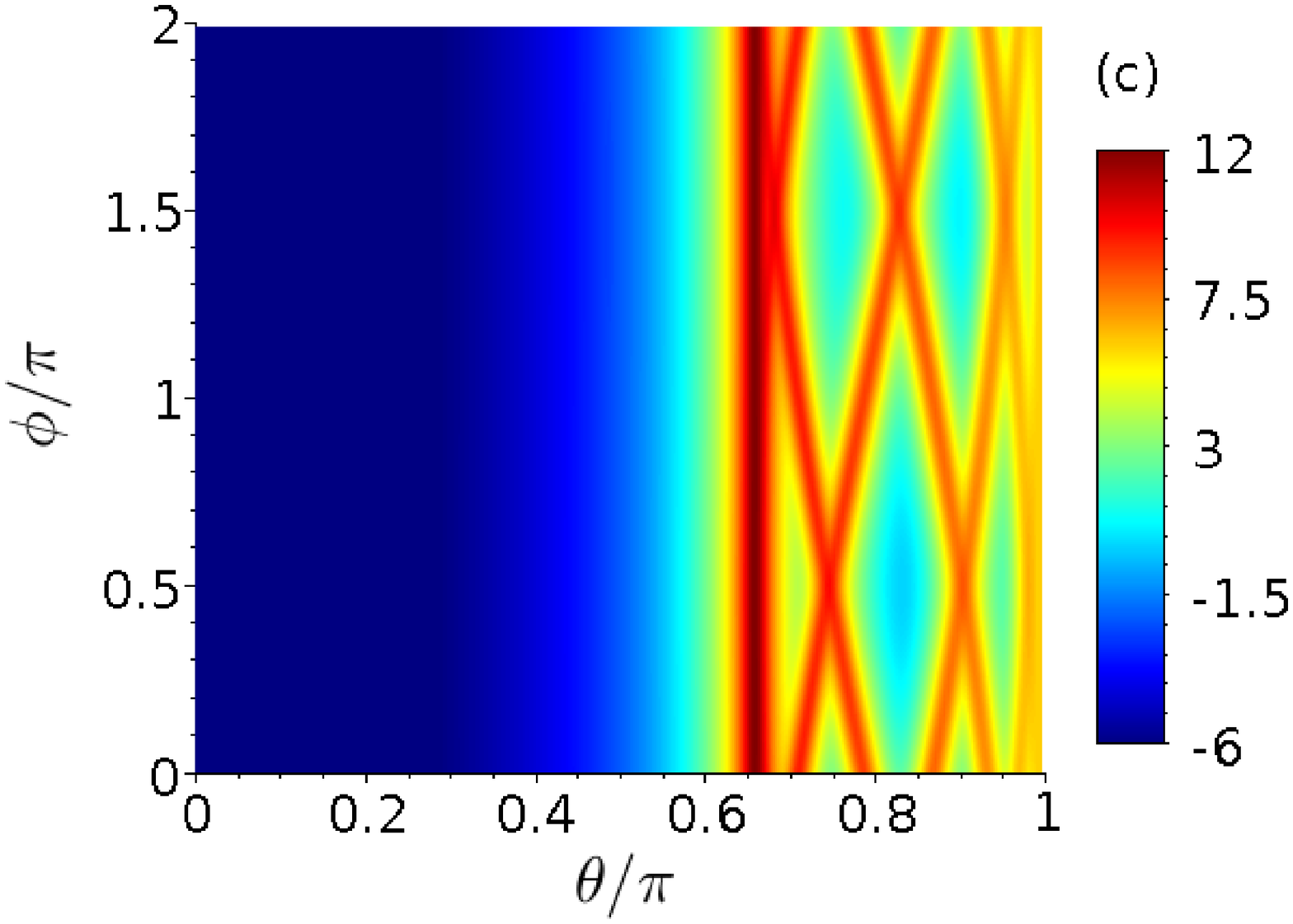}
\caption{(Color online) $\frac{dW}{d\Omega}$ for head on collision, $\tau=2$, $N_{\mathrm c}=10$. (a): $\gamma=10$; (b): $\gamma=17.7$; (b): $\gamma=30$.\label{f2}}
\end{figure}
Next we will study, for the same scattering geometry, the effect of the initial velocity of the electron on the angular distribution $dW/d\Omega$. The laser pulse was chosen with longer wings ($\tau=2$)  than previously and with a flat region of length $N_{\mathrm c}=10$.
In Fig. \ref{f2}(a) is represented $dW/d\Omega$ for $\gamma=10$; the figure is very similar to Fig. \ref{f1}(b). The bright line, which is the contribution of the constant part of the pulse, appears at the same value of $\theta=\Theta_0\approx0.32\pi$ as in Fig. \ref{f1}(b) since $\gamma$ and $\eta$ have the same values; the structure present in the region $\theta>\Theta_0$ is richer than in the previous case due to the fact that the wings of the envelope are longer, containing thus more oscillations of the carrier.  An interesting particular  situation met in a head-on collision is  that when  the initial velocity of the electron  compensates the forward drift caused by the laser field and as a consequence the bright maximum appears at $\Theta_0=\pi/2$. From Eq. (\ref{def-Z0}) the condition to be satisfied  for this is
\begin{equation}
\frac{e^2A_0^2}4+p_{0z}(n_{\mathrm L}\cdot p_0)=0\label{g0}
\end{equation}
and is equivalent in fact to the condition of cancellation of the third component of the ``dressed electron'' momentum
\begin{equation}
q_0=p_0+\frac{e^2A_0^2}{4(n_{\mathrm L}\cdot p_0)}n_{\mathrm L}\label{def-q}.
\end{equation}
For the intensity considered here the equation (\ref{g0}) leads to the solution $\gamma=17.7$ and the corresponding plot is represented in Fig. \ref{f2}(b). When the  electron energy increases further, the spectrum is compressed toward higher values of $\theta$; in Fig. \ref{f2}(c)  the case  $\gamma=30$ is considered,   the corresponding $\Theta_0$ being about $0.65\pi$.
\begin{figure}
\includegraphics[scale=0.24]{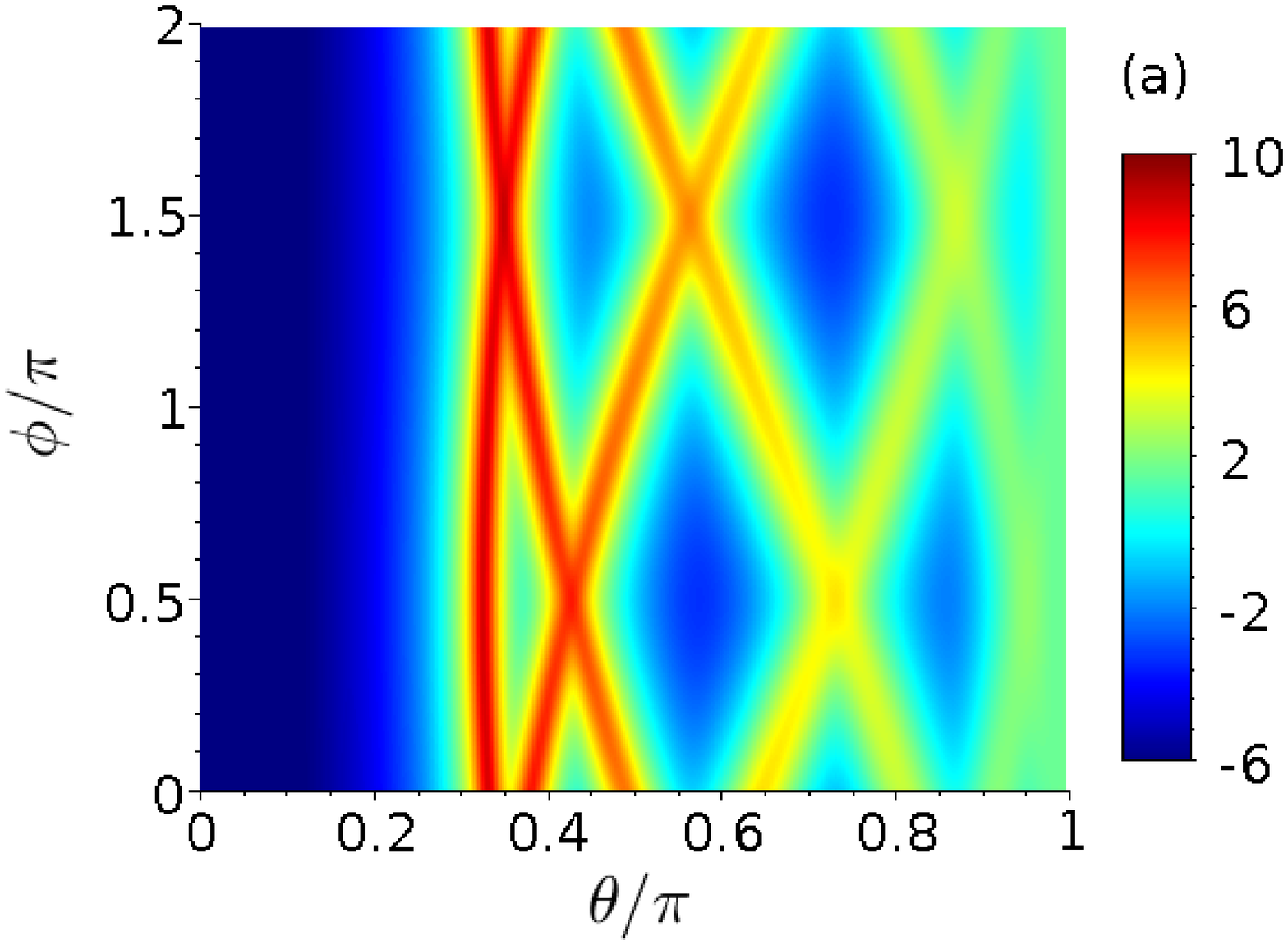}
\includegraphics[scale=0.24]{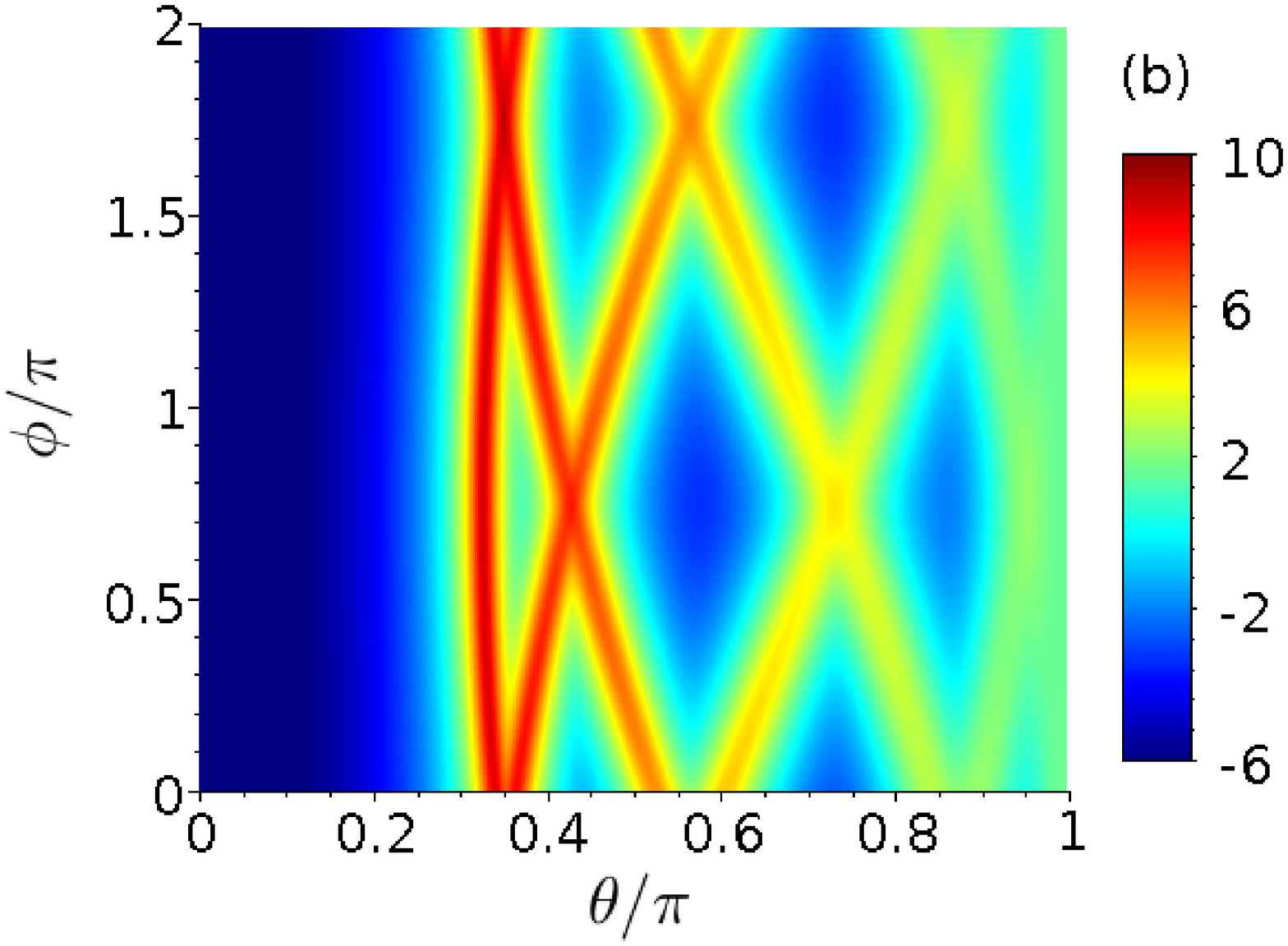}
\includegraphics[scale=0.24]{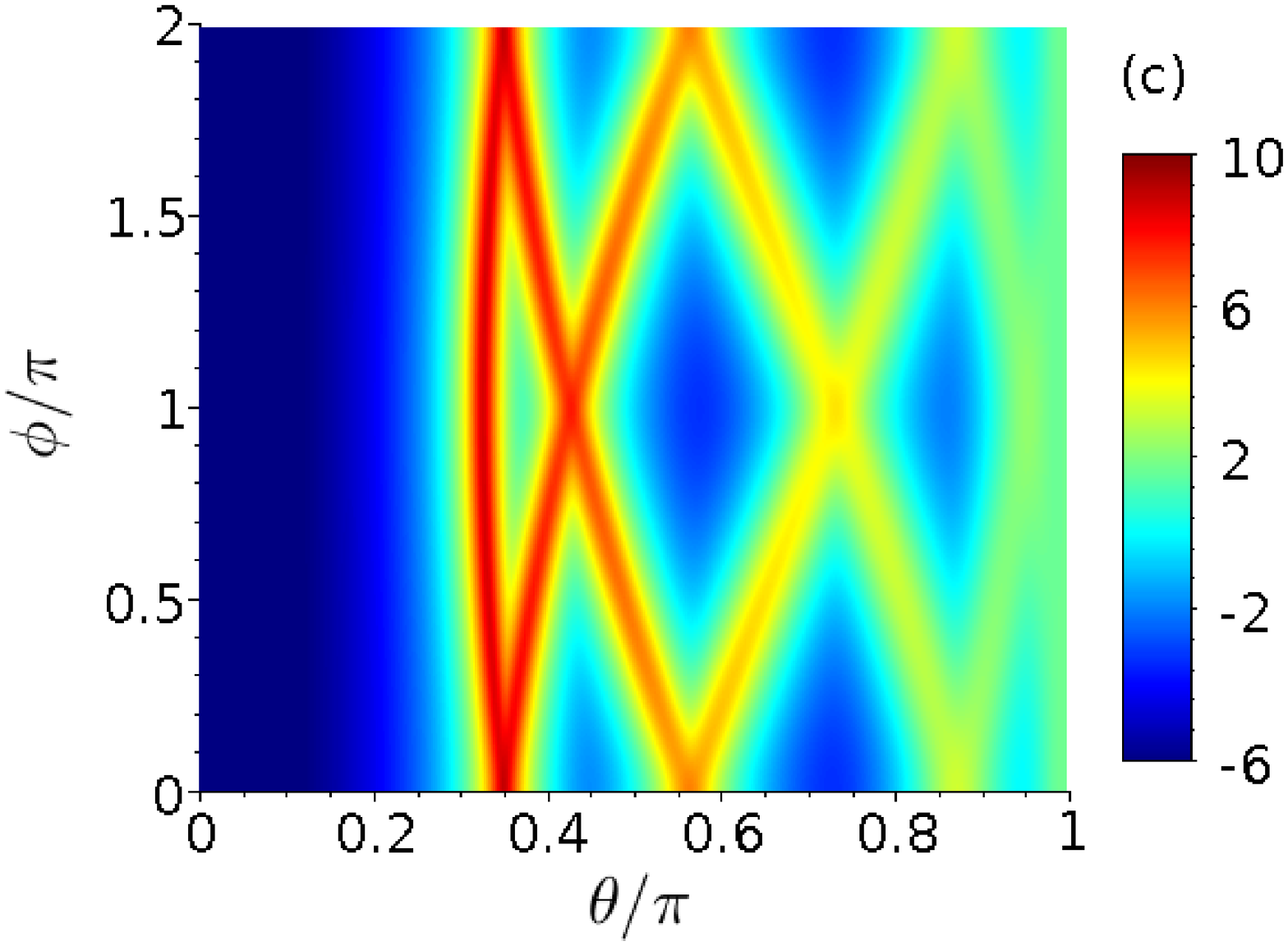}
\caption{(Color online)  $\frac{dW}{d\Omega}$ for head on collision, $\tau=1$, $N_{\mathrm c}=0$, $\gamma=10$ and for the vector potential (\ref{def-A-CEP}); (a): $\phi_0=0$, (b): $\phi_0=\pi/4$, (c): $\phi_0=\pi/2$ \label{f3}}
\end{figure}

From the previous results we have seen that there is a very strong connection between the angular distribution of the emitted radiation and the velocity of the electron, which, in its turn is determined by the shape of the laser pulse. It follows that it might be possible the reconstruction of the electromagnetic  pulse instead of field from the recorded angular distribution. Such an attempt is favored by the fact that the contribution of the radiation emitted during the interaction of the electron with the central part of the pulse, which could be much larger than the contribution of the beginning and the end  regions of the pulse, appear in different regions of the plane $(\theta,\phi)$, so they do not mask each other.  As an example, in Fig. \ref{f3} we present the effect of changing the relative phase of the envelope with respect to the carrier (CEP). In order to do this, we  use a modified vector potential, including a initial phase of the carrier,
 \begin{equation}
{\bf A}(\chi)=\frac{A_0}{\sqrt{2}}f(\chi)[{\bf e}_x\sin(k_{\mathrm L}\chi-\phi_0)+{\bf e}_y\cos(k_{\mathrm L}\chi-\phi_0)].\label{def-A-CEP}
\end{equation}
The parameters of the laser pulse are $\tau=2$, $N_{\mathrm c}=0$ and the Lorentz factor of the electron is $\gamma=10$. The scattering geometry is again head-on, such that the change of the initial phase $\phi_0$ is equivalent to a rotation around the $Oz$ axis. This property is illustrated in Fig. \ref{f3}; in the left graph (a) $\phi_0$ is 0, in the middle graph (b) $\phi_0=\pi/4$ and in the right graph (c) $\phi_0=\pi/2$; the three figures are identical, except for a shift of $\pi/4$ along the $\phi$ axis, in agreement with the previous discussion.

\begin{figure}
\includegraphics[scale=0.25]{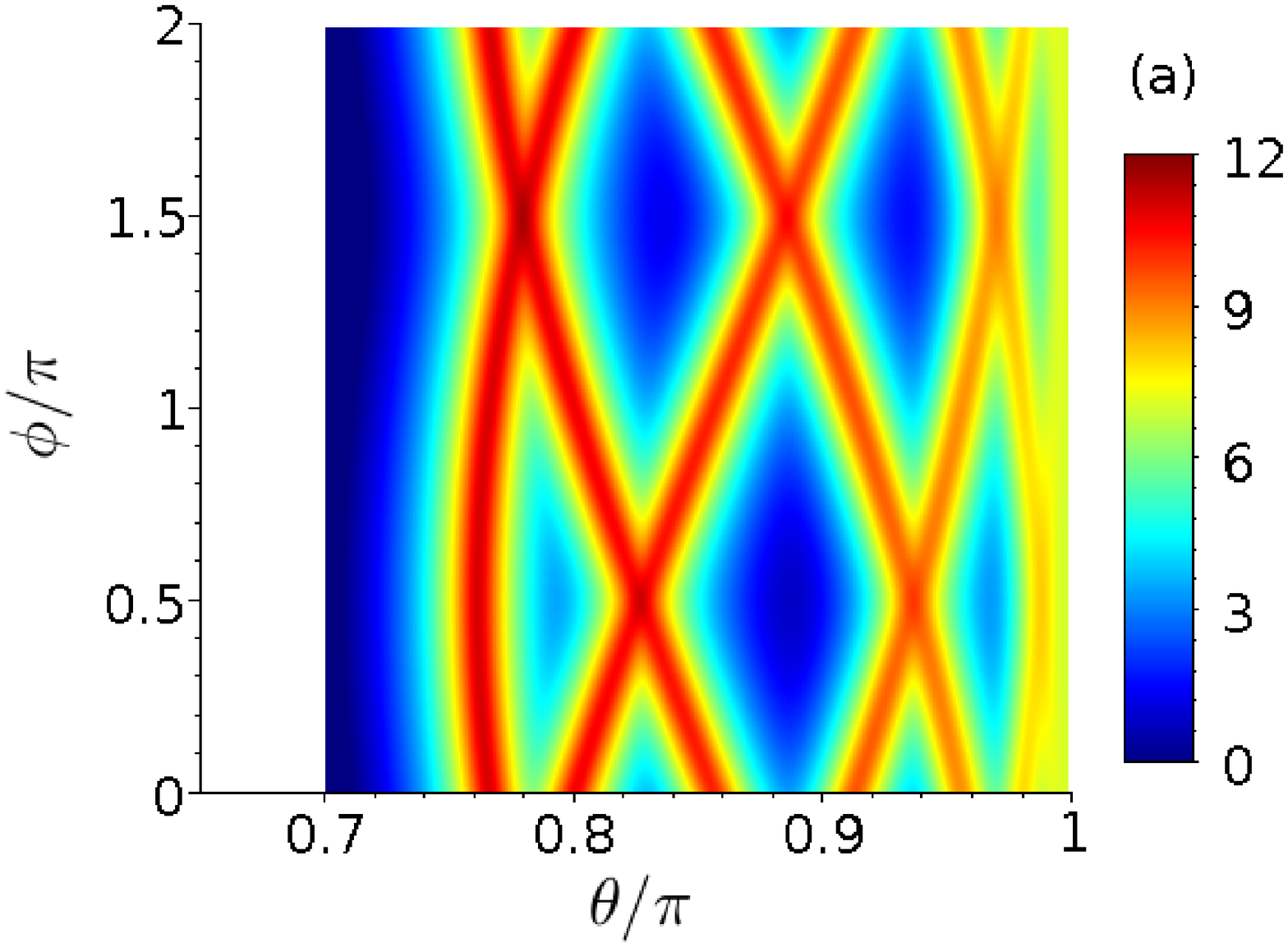} 
\includegraphics[scale=0.25]{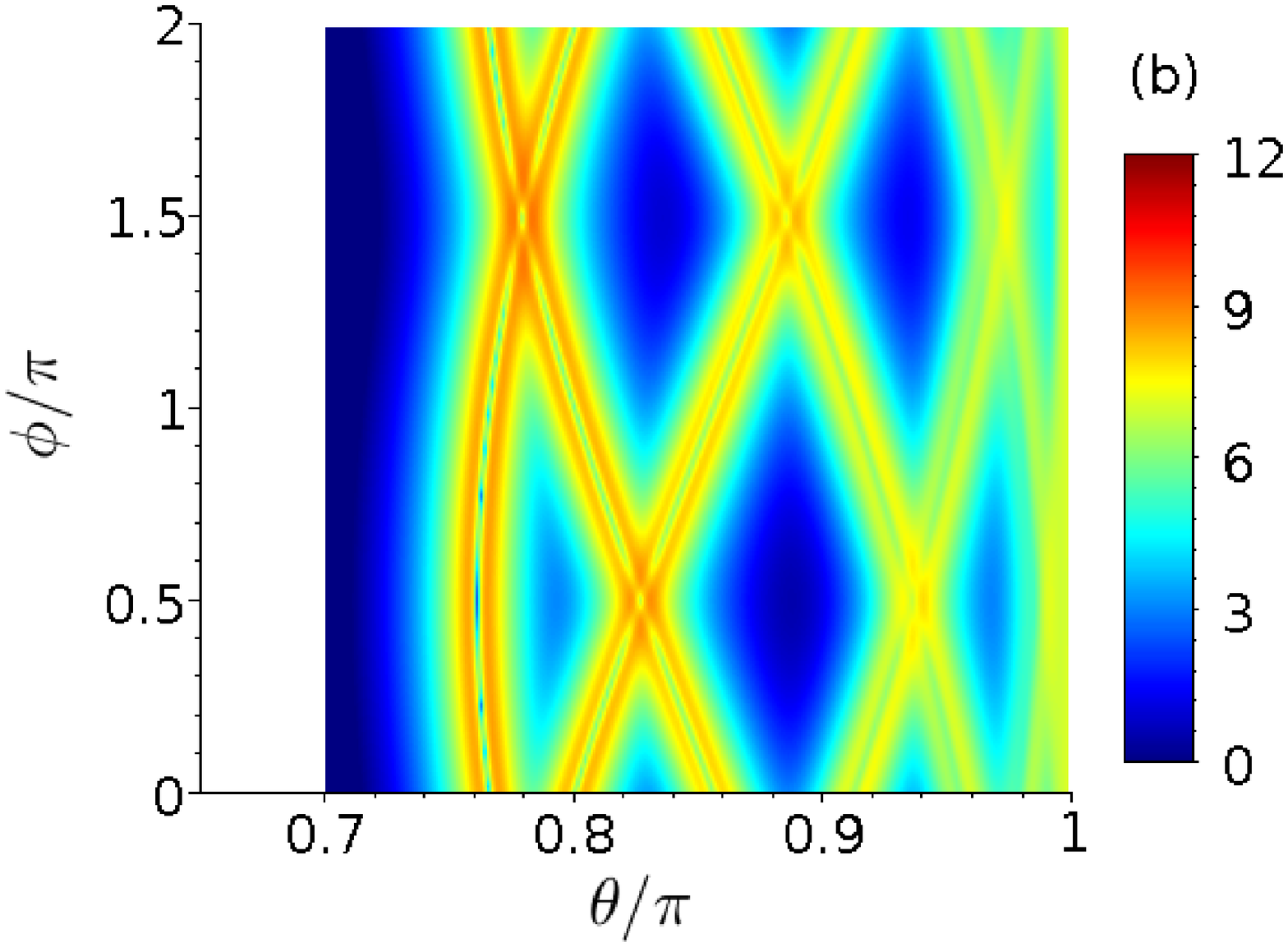}

\includegraphics[scale=0.25]{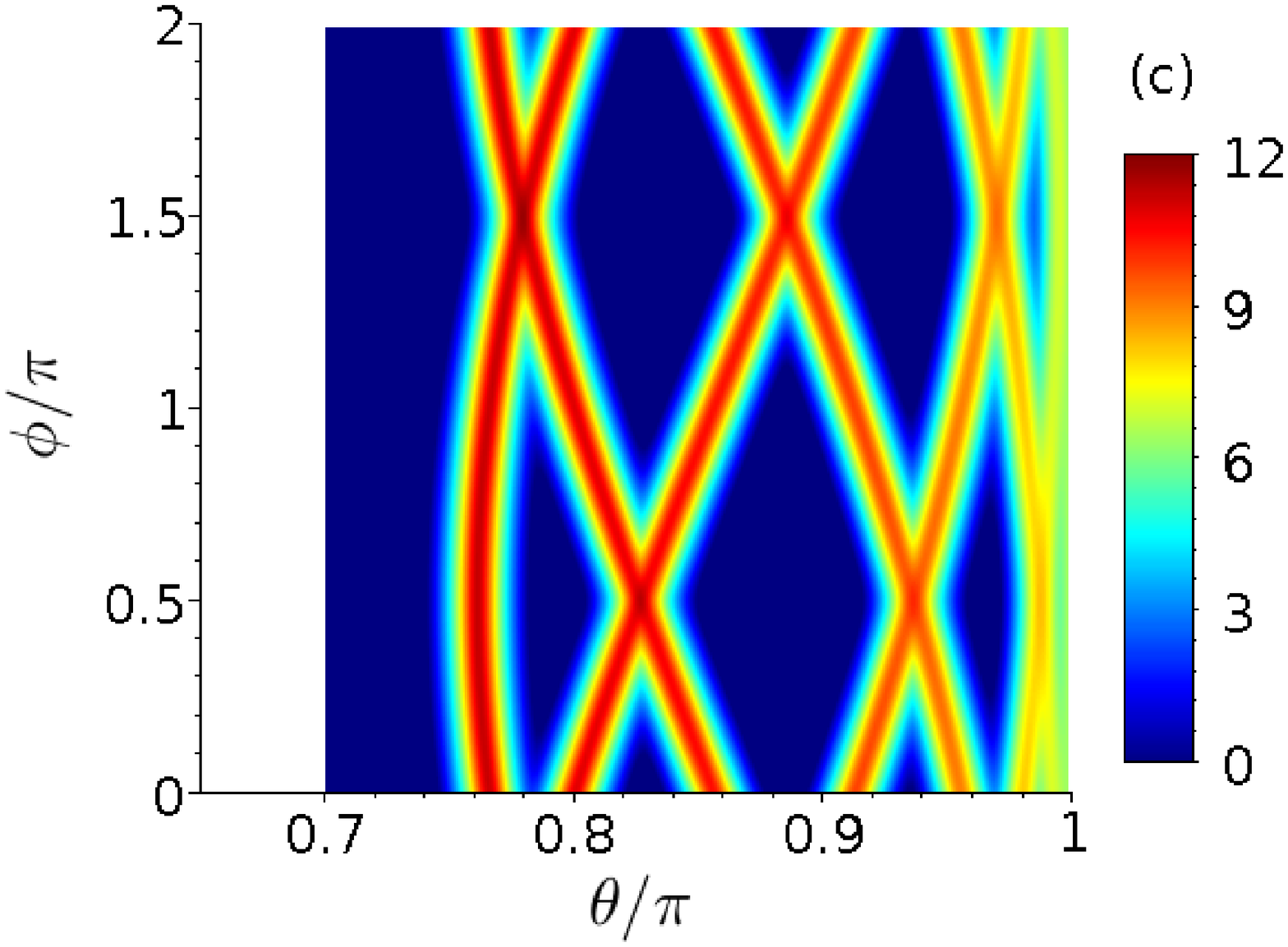} 
\includegraphics[scale=0.25]{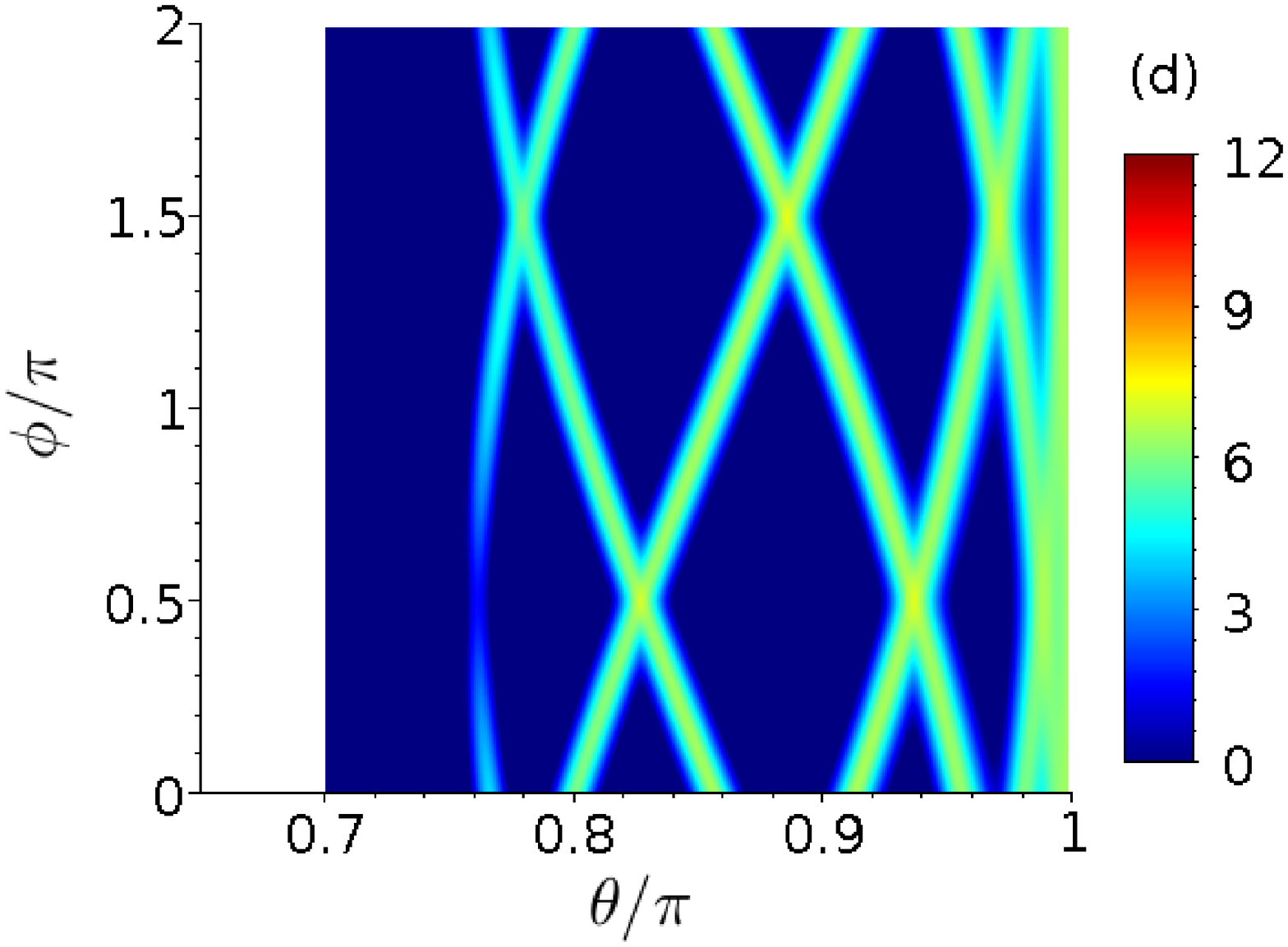}
\caption{(Color online) polarization analysis for head-on collision, $\tau=2$, $N_{\mathrm c}=0$, $\gamma=45$; (a): $dW_1/d\Omega$, (b): $dW_2/d\Omega$, (c): $d\widetilde W_1/d\Omega$, (d): $d\widetilde W_2/d\Omega$.\label{f4}}
\end{figure}

Next we present the results of the polarization analysis and the comparison between the exact classical formula and the approximation (\ref{dW-pol-app}) for a pulse with $\tau=2$, $N_{\mathrm c}=0$ and the electron Lorentz factor $\gamma=45$. In Fig. \ref{f4} are represented the contributions $dW_1/d\Omega$ (a), $dW_2/d\Omega$ (b) of the two polarization states defined in Eq. (\ref{dW-pol}), with the polarization vectors chosen as in Eq. (\ref{epsilon}), and the approximate results $d\widetilde W_1/d\Omega$ (c) and $d\widetilde W_2/d\Omega$ (d), calculated according to Eq. (\ref{dW-pol-app}).  As the initial electron energy is very large, the entire distribution is compressed in the region $\theta>0.75\pi$, but the shape is similar to the previous ones. One can see that $dW_1/d\Omega$ is larger than $dW_2/d\Omega$ by about two orders of magnitude; another difference is that while $dW_1/d\Omega$ has a maximum along the trajectory of $\hat{\bm{\beta}}$, $dW_2/d\Omega$ has a very sharp minimum surrounded by two adjacent maxima. The comparison of the exact results with the approximate ones shows that $d\widetilde W_1/d\Omega$ reproduces correctly the value along the velocity trajectory but the maxima are sharper than in the exact calculation. For the other polarization, although there is a good agreement of the values precisely along the trajectory of $\hat{\bm{\beta}}$, the adjacent maxima are not reproduced. The small value obtained for $d\widetilde W_2/d\Omega$ can be explained using its  expression (\ref{dW-pol-app}). As discussed in the previous section, for each observation direction ${\bf n}$ the contribution to the integral (\ref{dW-pol}) is given mainly by the values of $\chi$ satisfying the condition $\hat{\bm{\beta}}(\chi)\approx{\bf n}$. For the scattering geometry (\ref{head-on}) and the vector potential (\ref{def-A}) used here, one obtains for the component  of $\dot{\bm{\beta}}$ along the second polarization vector [Eq. (\ref{dot-beta-app})], calculated for ${\bf n}=\hat{\bm{\beta}}$
\begin{equation}
{\dot\beta_2\vphantom{\frac12}\vline}_{\,{\bf n}=\hat{\bm{\beta}}}=\frac c{|{\bm{\beta}}\times({\bm{\beta}}\times{\bf{n}}_{\mathrm L})|}\frac {e^2A_0^2f(\chi)\frac{df(\chi)}{d\chi}}{(1+F(\chi))^4}\left[F(\chi)-\frac{e^2{\bf A}^2(\chi)}{(n_{\mathrm L}\cdot p_1)^2}\frac{2+F(\chi)}{1+F(\chi)}\right];
\end{equation}
the derivative $df(\chi)/d\chi$ of the pulse envelope which appears as a global factor makes the entire result small, since the envelope is a slowly varying function  of $\chi$.

\subsection{The case of  the 90 degrees geometry}\label{ss-90}

 The second scattering geometry we  have investigated is  that of  a  collision at 90 degrees; the laser propagation direction is, as in the previous case, along the positive sense of the $Oz$ axis, and the initial velocity of the electron is chosen along the positive sense of the $Oy$ axis.

\begin{figure}
\includegraphics[scale=0.25]{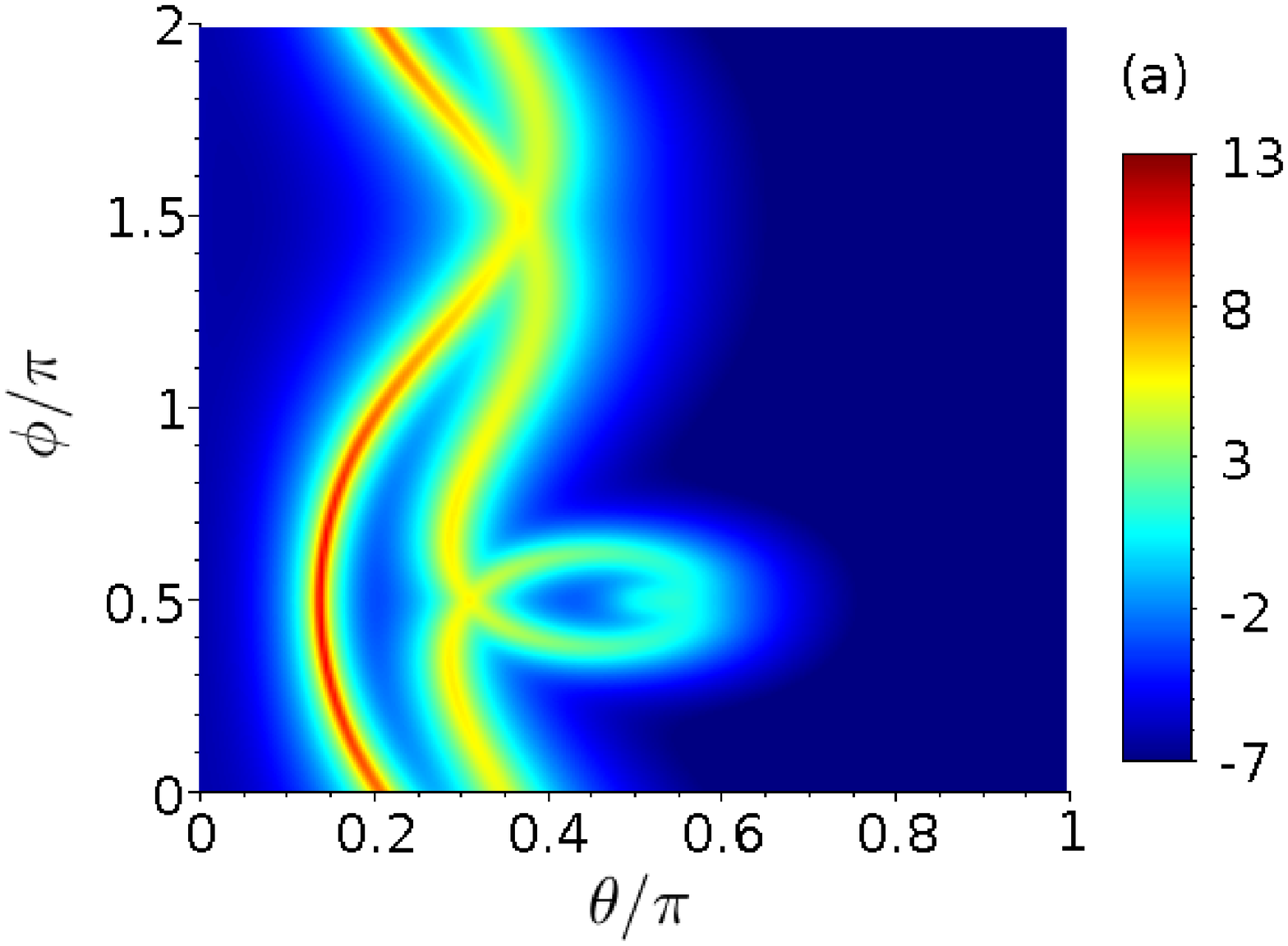} 
\includegraphics[scale=0.25]{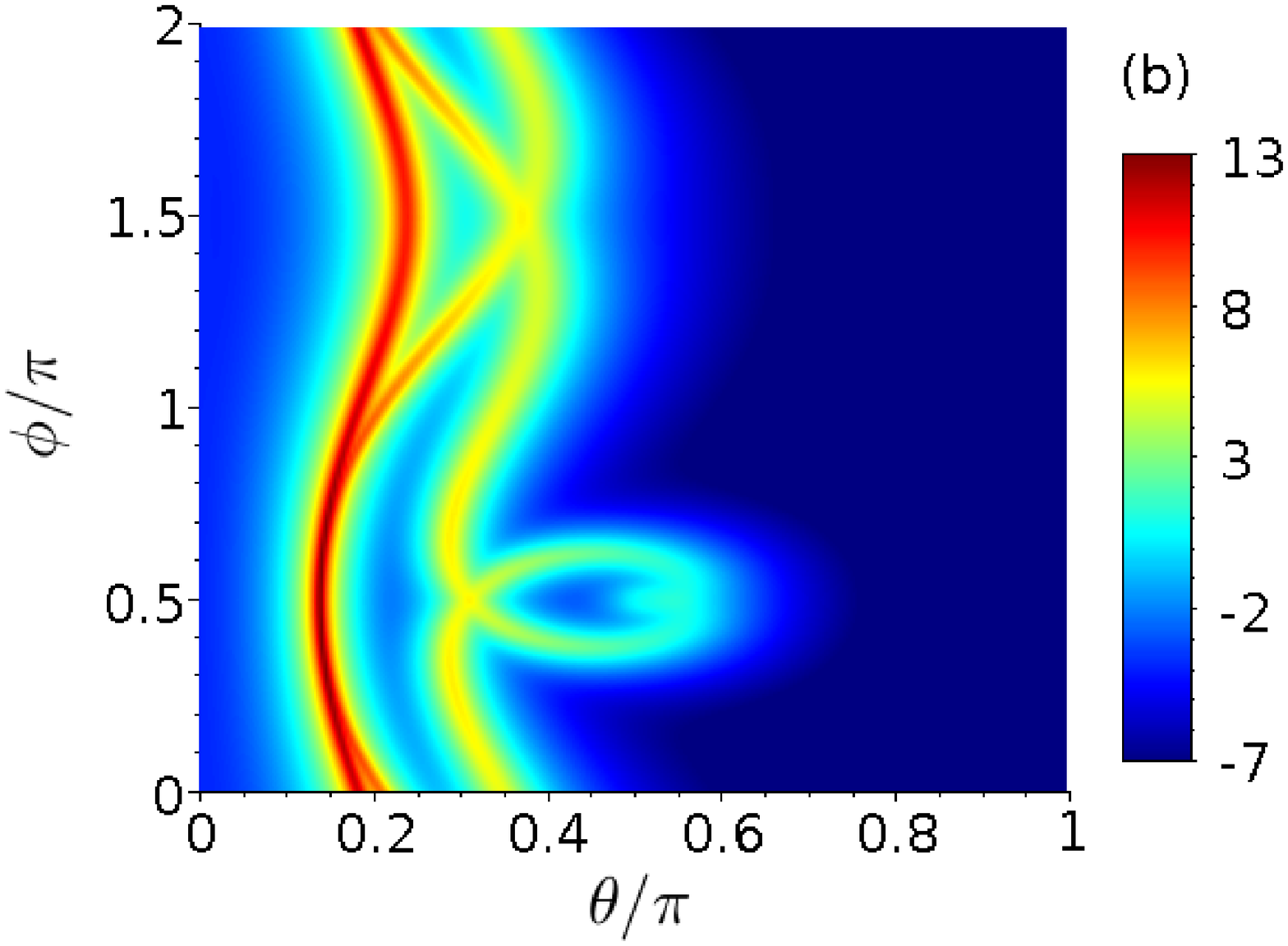}
\includegraphics[scale=0.19]{f5-c.eps}
\caption{(Color online)  (a): $\frac{dW}{d\Omega}$ for 90 degrees collision, $\tau=1$, $N_{\mathrm c}=0$, $\gamma=10$; (b): the same as (a), except for $N_{\mathrm c}=10$; (c) the trajectory of $\widehat{\bm{\beta}}$ in the plane $(\theta,\phi)$ in the same case as (b).\label{f5}}
\end{figure}

First, as in the case of head-on collision,  we shall compare the angular distribution $dW/d\Omega$  and  the velocity trajectory in the plane $(\theta,\phi)$. In Fig. \ref{f5}(a) is represented $dW/d\Omega$ for a pulse consisting in only   two wings  with $\tau=1$ and an  electron Lorentz factor $\gamma=10$; Fig. \ref{f5}(b) refers to the same case, except that the laser pulse has a constant region of length $N_{\mathrm c}=10$. Finally, in Fig. \ref{f5}(c) is represented the electron velocity trajectory, for the same conditions as in the case (b), with red line being represented  the contribution of the flat part of the pulse. Again, there is a very good  agreement between  the shape of $dW/d\Omega$ and the trajectory of $\widehat{\bm{\beta}}$, but there are differences with respect to the  case of head-on collision. The trajectory is more complicated,  the contribution of the flat region is not a straight line anymore, but it is still distinct from the contribution of the wings, and leads to a very intense sharp line in the spectrum.

\begin{figure}
\includegraphics[scale=0.24]{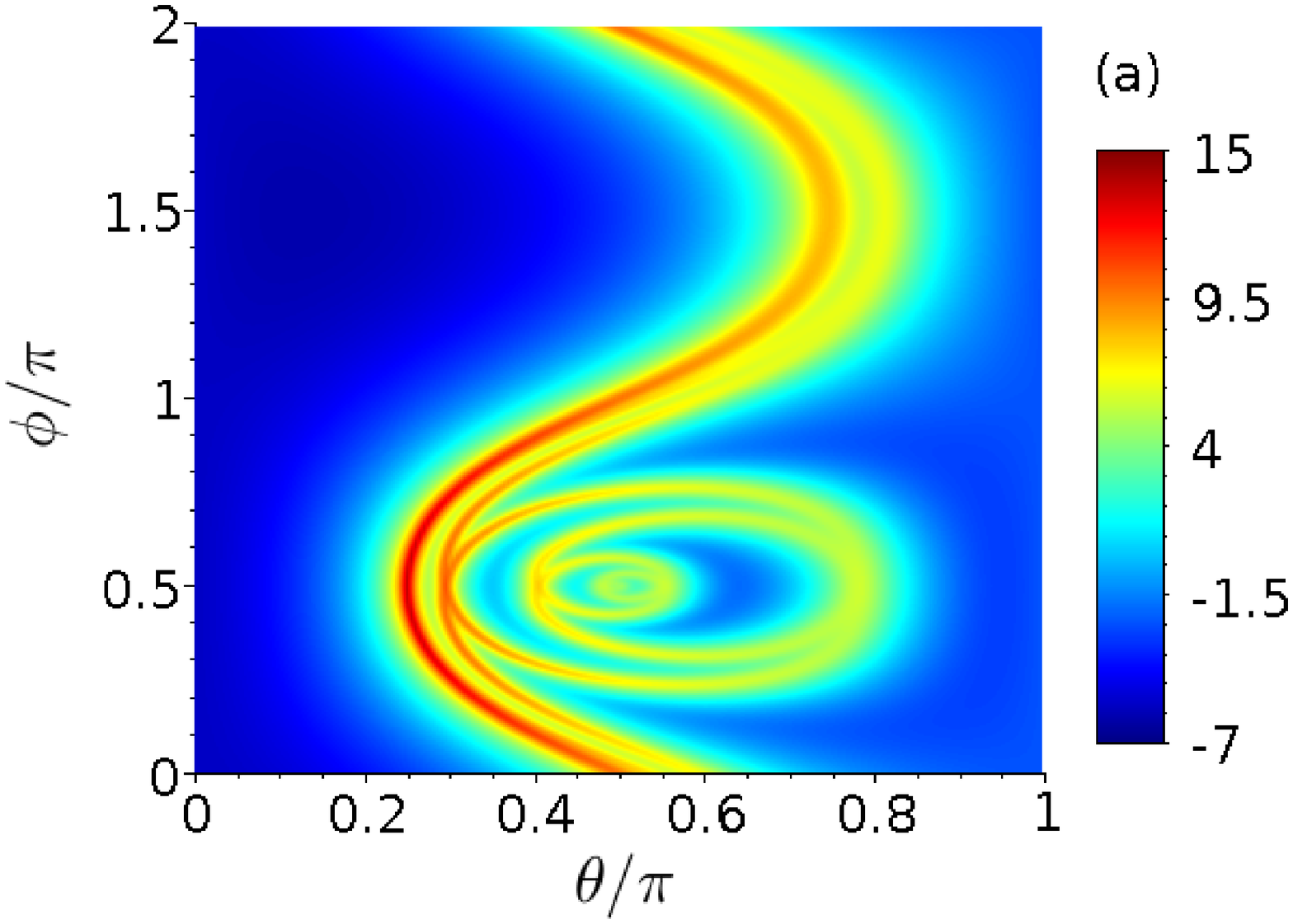} 
\includegraphics[scale=0.24]{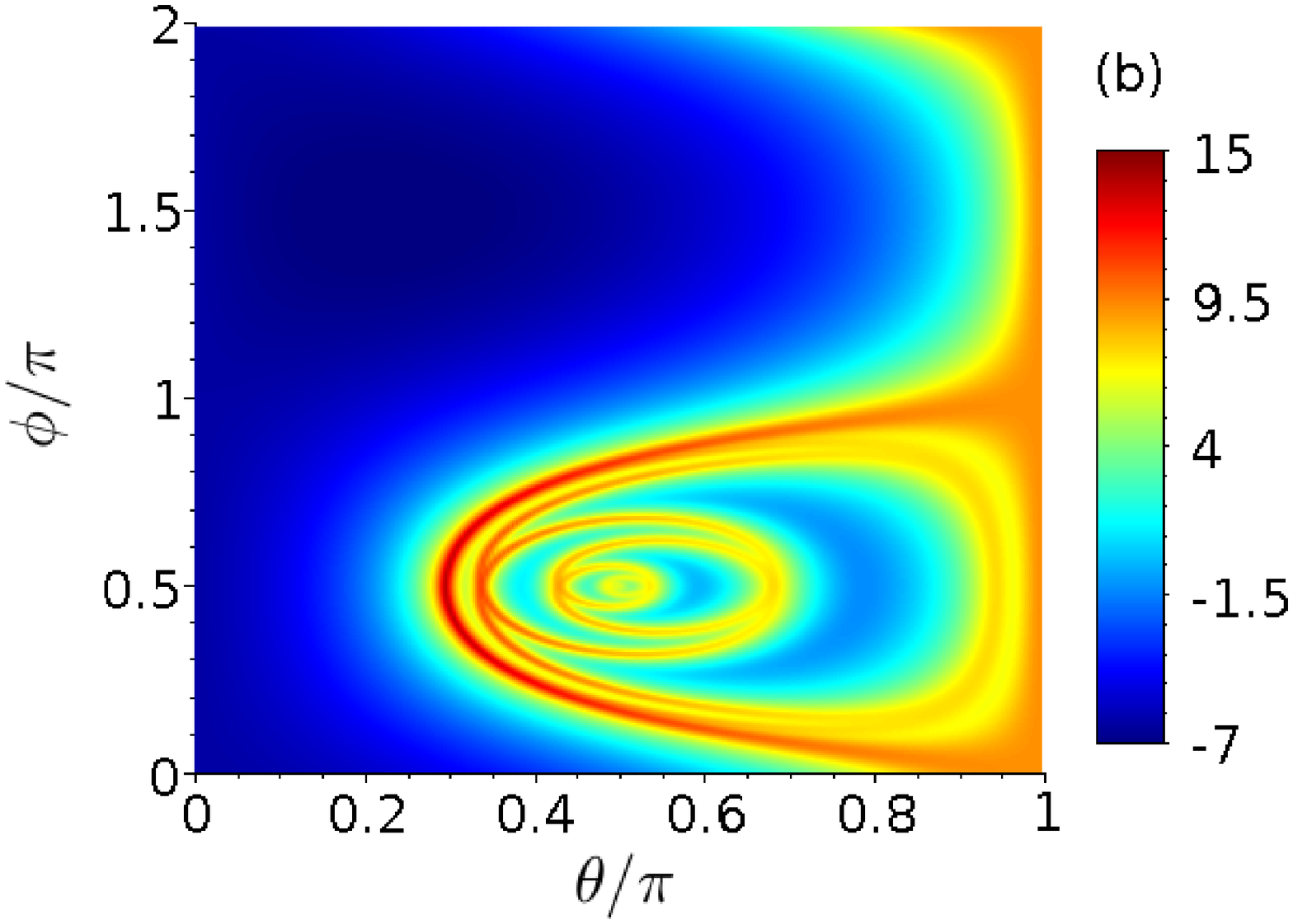}
\includegraphics[scale=0.24]{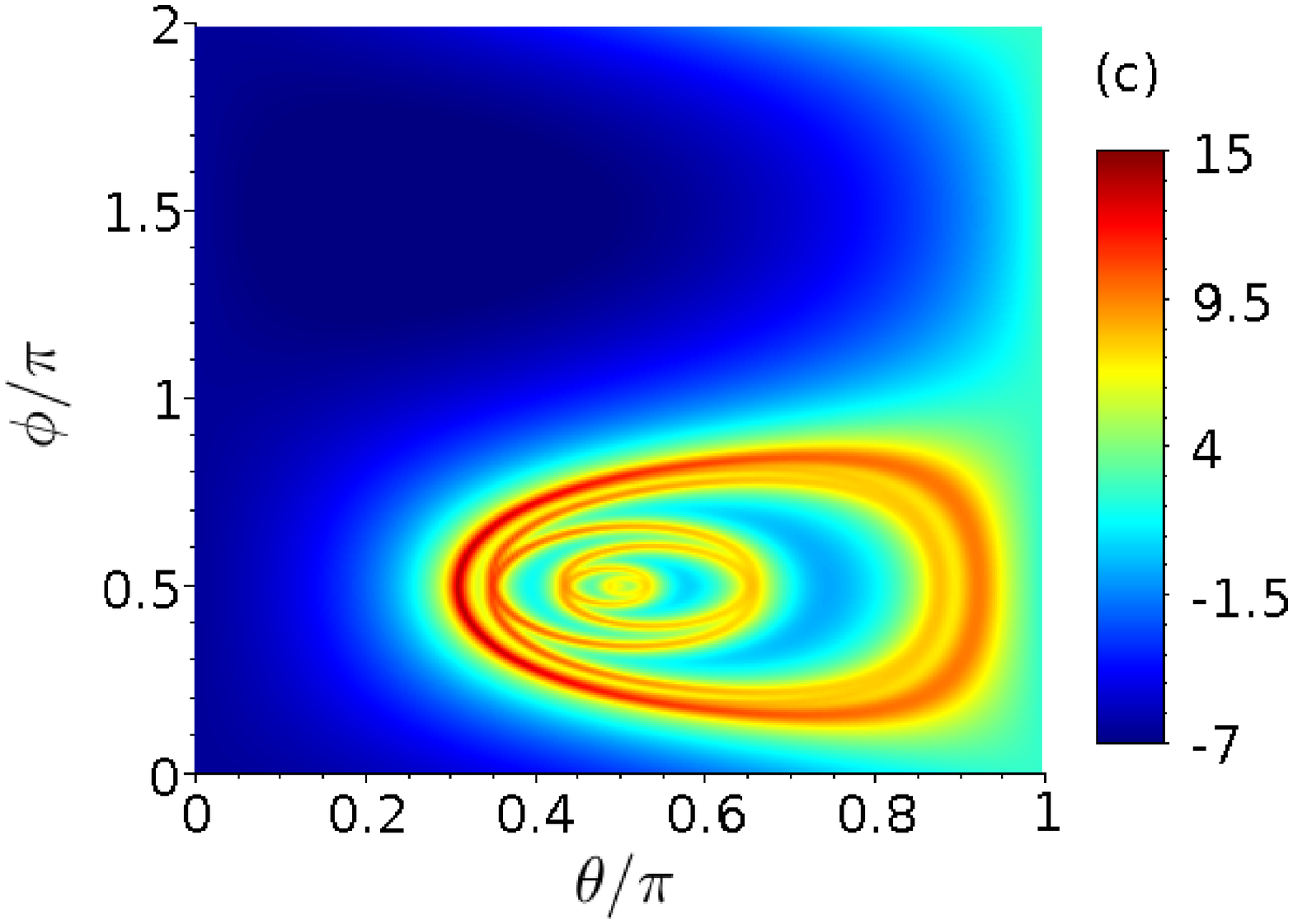}
\caption{(Color online) $\frac{dW}{d\Omega}$ for 90 degrees collision, $\tau=2$, $N_{\mathrm c}=10$. (a): $\gamma=25$; (b): $\gamma=35.37$; (b): $\gamma=40$.\label{f6}}
\end{figure}

In Fig. \ref{f6} we present the dependence of the angular distribution on the electron energy; one expects, as in the case of head-on collision, that for a very energetic electron the radiation to be emitted in a small cone whose axis coincides with the direction of the initial velocity, chosen along the $Oy$ axis. The laser pulse parameters are $\tau=2$ and $N_{\mathrm c}=10$. In Fig. \ref{f6}(a) the electron Lorentz factor is $\gamma=25$. In comparison with the case of   Fig. \ref{f5}(a), the emission takes place at angles $\theta$ larger, which is a consequence of the large electron initial energy; also, the graph has a richer structure, due to the fact that the pulse is longer.
From the velocity equations (\ref{eq-beta}) it follows  that for the geometry discussed here and for $p_{0y}\ge|e|A_0/\sqrt{2}$ the second component of velocity is positive at any moment, so in this case the emission should take place only in the semi-space $y>0$. The condition $p_{0y}=|e|A_0/\sqrt{2}$ leads to $\gamma=35.37$, and the corresponding angular distribution is represented in Fig. \ref{f6}(b); one can see that except for a small contribution near $\theta=\pi$ the emission takes place at angles $\phi\le\pi$. Figure \ref{f6}(c) corresponds to a large electron energy, $\gamma=40$. In this case, as it should be expected, the radiation is emitted within  a cone around the initial electron direction $\theta=\pi/2$, $\phi=\pi/2$.

\begin{figure}
\includegraphics[scale=0.25]{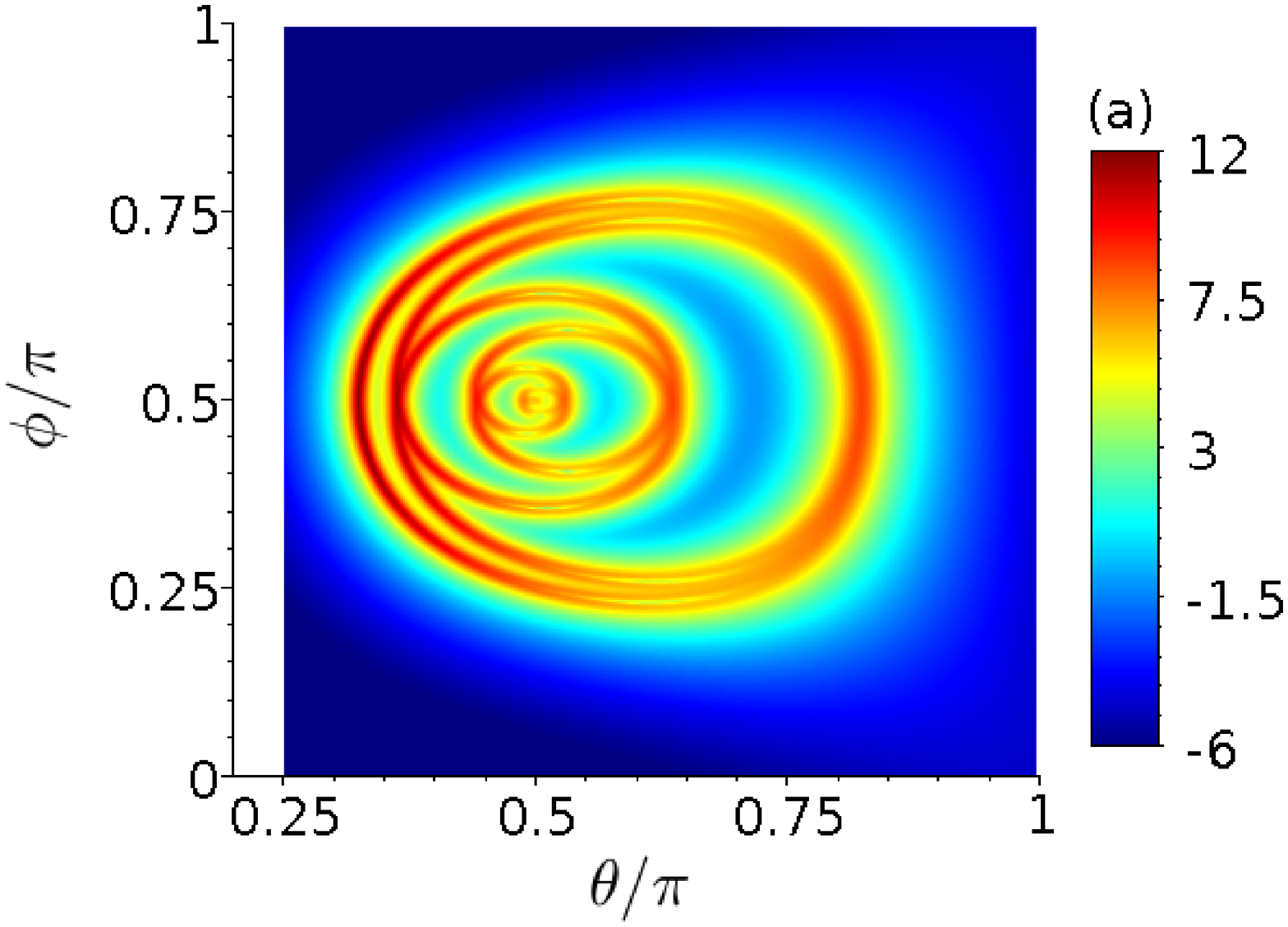} 
\includegraphics[scale=0.25]{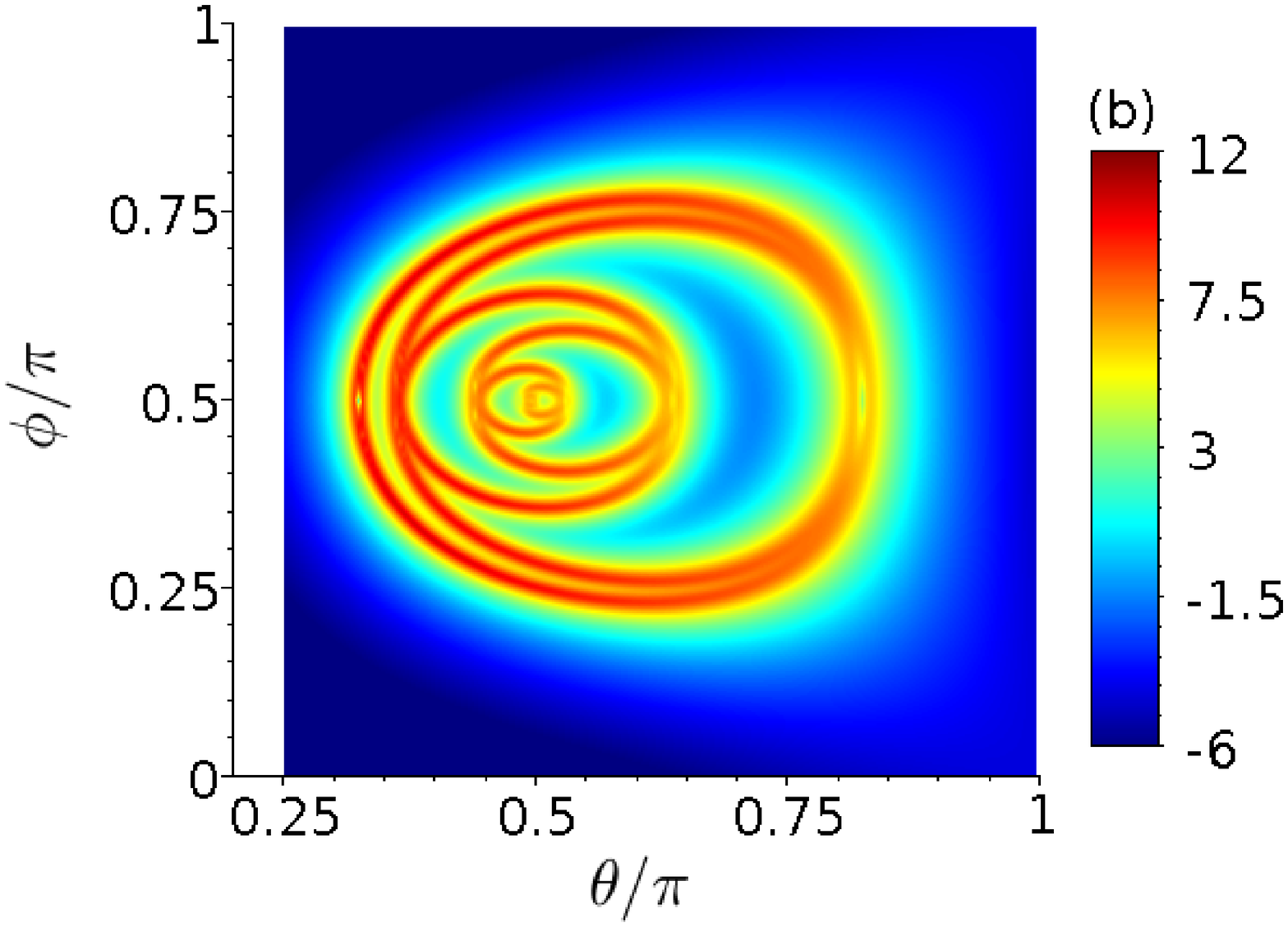}

\includegraphics[scale=0.25]{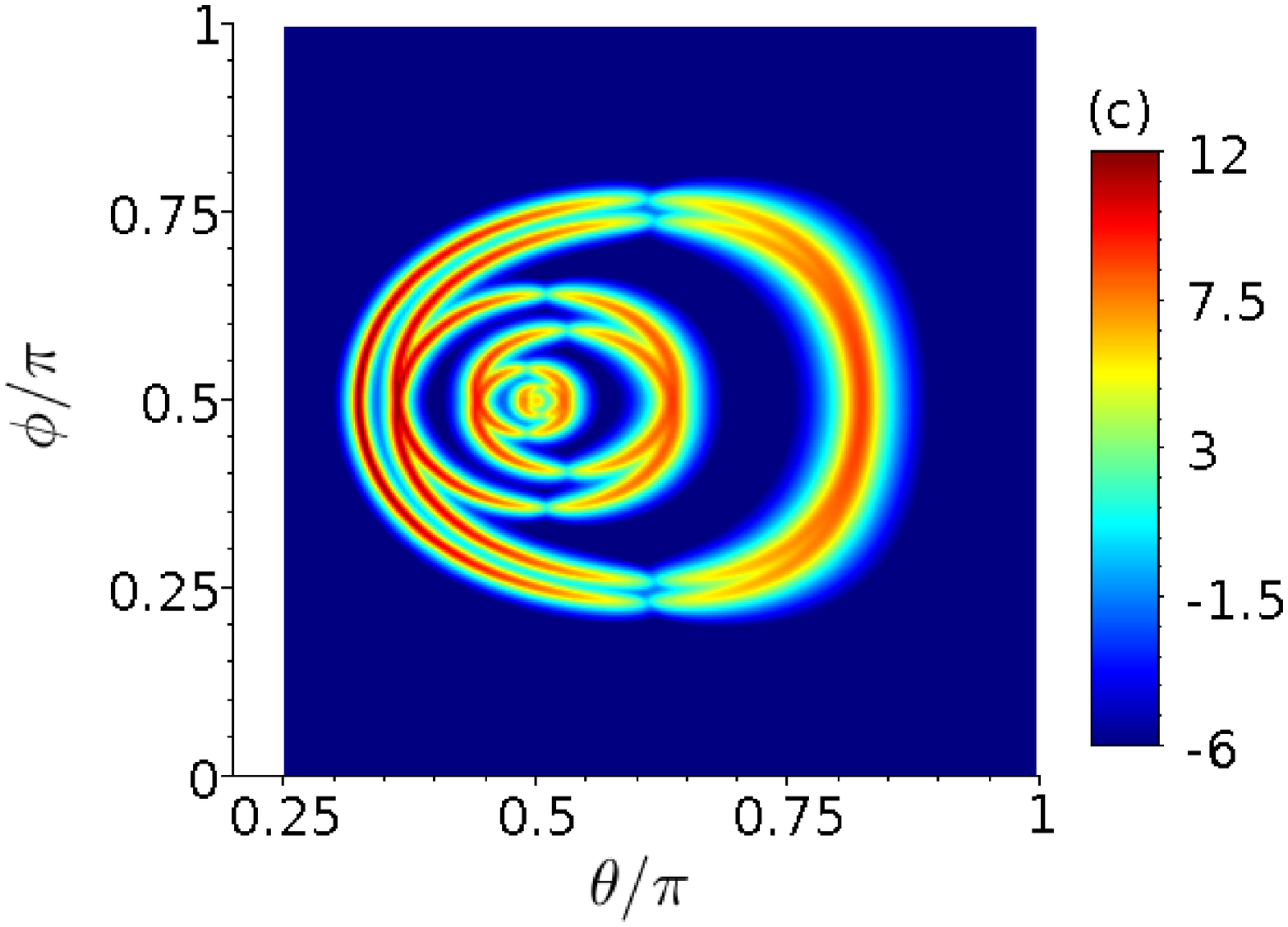} 
\includegraphics[scale=0.25]{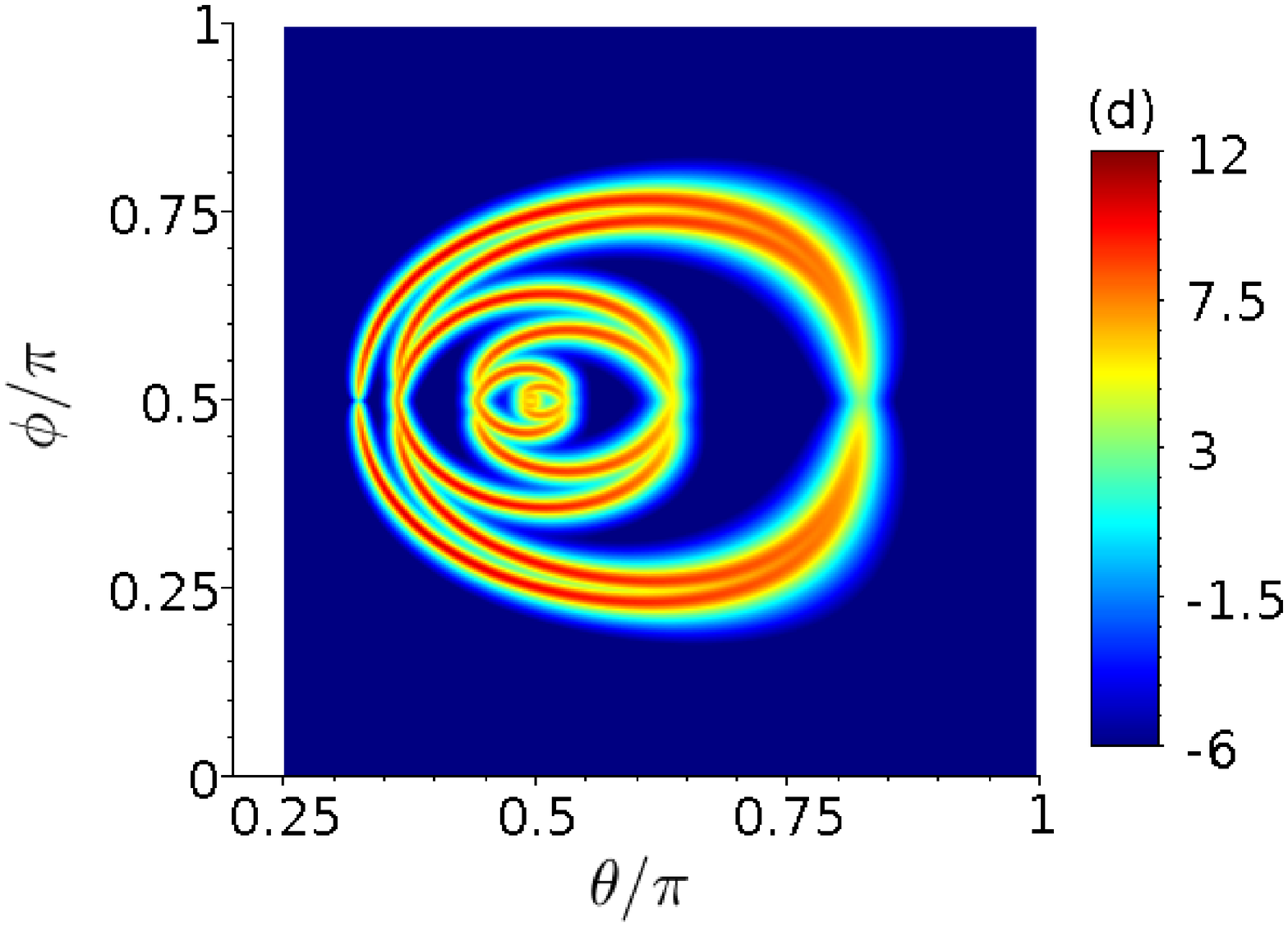}
\caption{(Color online) polarization analysis for 90 degrees collision, $\tau=2$, $N_{\mathrm c}=0$, $\gamma=45$; (a): $dW_1/d\Omega$, (b): $dW_2/d\Omega$, (c): $d\widetilde W_1/d\Omega$, (d): $d\widetilde W_2/d\Omega$.\label{f7}}
\end{figure}

Finally, for the same laser pulse ($\tau=2$, $N_{\mathrm c}=0$)  as in Fig. 6 and for a larger electron energy ($\gamma=45$) we present {an analysis of the} polarization.  The contributions  $dW_1/d\Omega$ and $dW_2/d\Omega$, defined in Eq. (\ref{dW-pol}), are represented in Figs. \ref{f7}(a) and \ref{f7}(b).  They have  similar shapes and, unlike in the case of head-on collision, the two are of the same order of magnitude. Also the range of angles in which the the radiation is emitted decreases further with respect to the case $\gamma=40$.   The approximate results calculated according to the formula (\ref{dW-pol-app}) are presented in Figs. \ref{f7}(c) and \ref{f7}(d); there is a good agreement   precisely  along the trajectory, but, as in the previous case the maxima are sharper than in the exact case. Even more, $d\widetilde W_1/d\Omega$ has very sharp minima near the positions along the $\widehat{\bm{\beta}}$ trajectory where the angle $\phi$ has a turning point; $d\widetilde W_2/d\Omega$ has very sharp minima at $\phi=\pi/2$. 
These minima can be understood using the approximate formula (\ref{dW-pol-app}). For the case of the first polarization the minima appear at the turning points of $\phi$; the correspondent condition for $\hat{\bm{\beta}}$ 
\begin{equation}
\frac{d}{d\chi}\arctan(\beta_y/\beta_x)=0
\end{equation}
can be written as
\begin{equation}
\frac{1+F(\chi)}{c}\frac{\dot\beta_y(\chi)\beta_x(\chi)-\dot\beta_x(\chi)\beta_y(\chi)}{\beta_x^2(\chi)+\beta_y^2(\chi)}=0
\end{equation}
By comparing the previous equation with the expression (\ref{dot-beta-app}) of the component of the acceleration along the polarization vector  ${\bm{\epsilon}}_1$ one can see that for those values of $\chi$ for which the velocity angle $\phi$ has a turning point, also ${\dot\beta_1\vline}_{\,{\bf n}=\hat{\bm{\beta}}}$ vanishes, leading to the minima in Fig. \ref{f7}(c). For the case of the second polarization the minima appear at $\phi=\pi/2$, i.e. when the observation direction is contained in the plane $Oyz$; in the framework of our approximation, the main contribution to the integral (\ref{dW-pol-app}) is given by those values of $\chi$ for which $\beta_x=0$. Using this condition in expression (\ref{dot-beta-app}) of the component of acceleration  along the ${\bm{\epsilon}}_2$ one obtains, as in the case of head-on collision,  that ${\dot\beta_2\vline}_{\,{\bf n}=\hat{\bm{\beta}}}$ is proportional to the derivative of the pulse envelope which explains the minima in Fig. \ref{f7}(d). For both polarizations  there is a correspondent of those minima as minima surrounded by sharp maxima in the graph of the exact results.

\subsection{Scaling laws}
Here we shall present two scaling laws for the angular distribution of the emitted radiation.

The first of them  refers to the dependence of the angular distribution on the laser central frequency $\omega_{\mathrm L}$; in a previous paper \cite{BF-u} it was shown that for envelopes depending only on the product $k_{\mathrm L}\chi$ the double differential distribution $d^2W/d\omega d\Omega$ is a function of $\omega/\omega_{\mathrm L}$ only
\begin{equation}
\frac{d^2W}{d\omega d\Omega}=f(\frac{\omega}{\omega_{\mathrm L}});
\end{equation} 
as a consequence  the angular distribution 
\begin{equation}
\frac{dW}{d\Omega}=\int\limits_0^{\infty}d\omega\frac{d^2W}{d\omega d\Omega}=\int\limits_0^{\infty}d\omega f(\frac{\omega}{\omega_{\mathrm L}})=\omega_{\mathrm L}\int\limits_0^{\infty}d\tilde\omega f(\tilde\omega)
\end{equation}
 is proportional to $\omega_{\mathrm L}$ so a change of the central frequency $\omega_{\mathrm L}$ leads only to the scaling of the numerical values of $dW/d\Omega$.

 The second scaling law which we  present    here is valid in the limit of high energy and refers to the {\it shape} of the angular distribution.  As we have previously seen, for the case of large  values of $\eta$ and $\gamma$  the angular distribution $dW/d\Omega$ has sharp maxima in the plane $(\theta,\phi)$ along the trajectory covered by the velocity unit vector during the interaction of the electron with the laser pulse. From the expression of the velocity (\ref{eq-beta}) one can see that in the high energy limit $\gamma\gg1$,  $\widehat{\bm{\beta}}(\chi)$ depends only on the ratio $\eta/\gamma$; as a consequence the {\it shape} of the angular distribution $dW/d\Omega$ presented  in our figures  for $\eta=50$ and  a given value of the electron energy should be identical to that obtained for other values of the parameter $\eta$, assuming that the Lorentz factor $\gamma$ is scaled accordingly.

\section{Conclusions}\label{s-conclusions}

We have investigated within the framework of   classical electrodynamics   the radiation emitted by an electron interacting with a very intense laser pulse ($\eta=50$), with fixed direction of propagation and finite  length.  We have considered initial electron energies in the range $10< \gamma<45$ and two scattering configurations: head-on geometry and 90 degrees geometry. We have shown that in all cases the angular distribution $dW/d\Omega$, represented as a function of the observation direction angles  $\theta$ and $\phi$, consists in a set of very sharp maxima whose distribution in the plane $(\theta,\phi)$ is identical to the trajectory followed by the unit vector of the particle velocity, $\widehat{\bm{\beta}}$. For large energies of the incident electron ($\gamma>18$ in the case of head-on collision and $\gamma>36$ for 90 degrees collision) the scattered radiation is emitted within a cone whose opening angle decreases with increasing of $\gamma$. We have presented the results of the polarization analysis for the two geometries, and in both cases we have compared the results with a high energy approximation, with the conclusion that the approximation reproduces at the qualitative level the exact calculation. Finally, for the case of head-on collision we have discussed the effect of the change of the relative phase between the pulse envelope and carrier and we have presented a numerical example.  The two scaling laws mentioned  in Sect. \ref{s-numerical}, namely the proportionality of the angular distribution with $\,\omega_L\,$ and the dependence of its shape only on the ratio $\,\gamma/\eta\,$ for $\gamma \gg 1$,  allow  the use of the results presented  in this paper for other values of the laser frequency and intensity.

\acknowledgments
The authors are grateful to Viorica Florescu for her guidance during the preparation of this paper and constant encouragements. This work was supported by CNCSIS-UEFISCSU, project number 488 PNII-IDEI 1909/2008.

\end{document}